\documentclass[12pt,preprint]{aastex}
\newcommand\etal{{\it et al.~}}
\newcommand\beq{\begin{equation}}
\newcommand\eeq{\end{equation}}
\newcommand\beqar{\begin{eqnarray}}
\newcommand\eeqar{\end{eqnarray}}

\newcommand\bvec[1]{\hbox{\boldmath${#1}$}}

\newcommand{\vef}{v_{{\rm e},\phi}}
\newcommand{\vif}{v_{{\rm i},\phi}}
\newcommand{\vgof}{v_{{\rm g0},\phi}}
\newcommand{\vgpf}{v_{{\rm g+},\phi}}
\newcommand{\vgmf}{v_{{\rm g-},\phi}}
\newcommand{\vi}{v_{\rm i}}
\newcommand{\ve}{v_{\rm e}}
\newcommand{\vn}{v_{\rm n}}
\newcommand{\vgo}{v_{\rm g0}}
\newcommand{\vgp}{v_{\rm g+}}
\newcommand{\vgm}{v_{\rm g-}}

\newcommand{\tz}{\tau_0}
\newcommand{\tp}{\tau_+}
\newcommand{\tm}{\tau_-}

\newcommand{\tne}{\tau_{\rm ne}}
\newcommand{\tngo}{\tau_{\rm ng0}}
\newcommand{\tni}{\tau_{\rm ni}}
\newcommand{\tngp}{\tau_{\rm ng+}}
\newcommand{\tngm}{\tau_{\rm ng-}}
\newcommand{\tin}{\tau_{\rm in}}

\newcommand{\tgn}{\tau_{\rm gn}}
\newcommand{\rn}{\rho_{\rm n}}

\newcommand{\ri}{\rho_{\rm i}}
\newcommand{\rgo}{\rho_{\rm g0}}
\newcommand{\rgp}{\rho_{\rm g+}}
\newcommand{\rgm}{\rho_{\rm g-}}
\newcommand{\ngm}{n_{\rm g-}}

\newcommand{\oitin}{\omega_{\rm i} \tau_{\rm in}}
\newcommand{\ogtgn}{\omega_{\rm g}\tau_{\rm gn}}

\newcommand{\rmm}{\frac{\rgo}{\rgm}\frac{\tgn}{\tm}}
\newcommand{\rpp}{\frac{\rgo}{\rgp}\frac{\tgn}{\tp}}

\begin{document}

\title{MAGNETICALLY-CONTROLLED SPASMODIC ACCRETION DURING STAR FORMATION: 
\\I. FORMULATION OF THE PROBLEM \\ AND METHOD OF SOLUTION} 

\author{Konstantinos Tassis \& Telemachos Ch. Mouschovias}

\affil{Departments of Physics and Astronomy \\
University of Illinois at Urbana-Champaign, 1002 W. Green Street, Urbana, IL 61801}

\begin{abstract}

We formulate the problem of the late accretion phase
of the evolution of an isothermal magnetic disk surrounding a forming
star. The evolution is described by the six-fluid MHD equations, accounting 
for the presence of neutrals, atomic \& molecular ions, electrons, and   
neutral, positively and negatively charged grains. 
Only the electron fluid is assumed to be attached to the 
magnetic field, in order to investigate the effect of the detachment of 
the ions from the magnetic field lines that begins at densities as low as  
$10^{8}$ $\rm{cm}^{-3}$. 
The ``central sink approximation'' is used to circumvent the 
problem of describing the evolution inside the opaque central region for
densities greater than $10^{11}$ $\rm{cm}^{-3}$. In this way, the structure 
and evolution of the isothermal disk surrounding the forming star can be 
studied at late times without having to implement the numerically 
costly radiative transfer required by the physics of the 
opaque core. The mass and magnetic flux accumulating  
in the forming star are calculated, as are their effects on the structure \& 
evolution of the surrounding disk.

The numerical method of solution first uses an adaptive grid and later, after 
a central region a few AUs in radius becomes opaque, switches to a stationary but 
nonuniform grid with a central sink cell. It also involves
an implicit time integrator; 
an advective difference scheme that possesses the transportive property; 
a second-order difference approximation of forces inside a cell; 
an integral approximation of the gravitational and magnetic fields; 
and tensor artificial viscosity that permits an accurate investigation of the 
formation and evolution of shocks in the neutral fluid.
\end{abstract}


\section{INTRODUCTION}

Interstellar molecular clouds are known to be the sites of star
formation. Typical sizes range from 1 to 5 pc,
masses from a few tens to $10^5$ ${\rm M_{\odot}}$,
and mean densities from $10^2$ to $10^3$ ${\rm cm^{-3}}$. Denser
fragments (or cores) are formed within individual clouds. Approximately 1\% of the
mass of a cloud is in the form of dust particles (grains). These
clouds are cold, with temperatures $\approx 10$ K, while their spectral lines have
Doppler-broadened  linewidths that suggest supersonic (but
subAlf\'{e}nic) internal
motions. Polarization maps \citep{Hildeb,Rao,Lai} reveal large-scale,
ordered magnetic fields in the plane of the sky. OH Zeeman
observations \citep{Crutetal,Crut94} find magnetic field strengths
between $10$ and $150$ ${\rm \mu G}$. 

 Observations also reveal a degree of ionization $\chi_i \gtrsim 10^{-4}$ in cloud
envelopes but $\lesssim 10^{-7}$ in dense cores \citep{Caselli, Williams,
  Bergin}. In 
the deep interior of molecular clouds, cosmic-ray ionization 
dominates other mechanisms, but in cloud envelopes UV ionization is also
important. A fraction of the dust grains becomes charged (negatively and/or 
positively depending on the density), and the grains thereby
play an important role in the fragmentation of molecular clouds and star
formation.  

A survey of protostars in the Orion Nebula \cite{Beckwith} found that almost
half of the star 
forming cores are surrounded by dusty disks. In addition, the
geometry of protostellar fragments is often disklike over a wide range of
lengthscales (Lay \etal 1994; Sargent \etal 1988; Kaifu \etal 1984).
Nevertheless molecular clouds and their cores rarely exhibit 
rotation significantly higher than that of the background medium, and when
they do, their angular velocities 
(typical core rotational rates are $\Omega \lesssim 3 \times 10^{-14}$ ${\rm
  sec^{-1}}$) imply centrifugal forces much too small to impose a
disklike geometry through rotational support perpendicular to the axis of
rotation \cite{Saito}. These angular velocities are much smaller than those
expected from angular momentum conservation from 
the initial galactic rotation (initial angular velocity $\Omega_0 \simeq
10^{-15}$ ${\rm s^{-1}}$ and density $n \simeq 1$ ${\rm cm^{-3}}$)
\cite{Goldsmith}.  
If angular momentum were to be conserved, centrifugal forces would
not allow even the formation of interstellar clouds (Mouschovias 1991,
eq. [2]). This
is one way of stating the `` angular momentum problem'' of star formation. 

The presence of magnetic fields in interstellar molecular clouds poses
another serious problem for star formation. The magnetic flux of an
interstellar blob of mass comparable to a stellar mass is typically between two
and five orders of magnitude greater than observed fluxes of even strongly
magnetic young stars (Basri, Marcy \& Valenti 1992; Wade \etal
1998). This suggests that flux loss takes place at some stage
during star formation.

Despite the fact that the critical mass for collapse of a pressure-bounded,
isothermal, nonmagnetic, nonrotating spherical cloud is only $5.8$ ${ \rm
  M_{\odot}}$  at
density $n = 10^3$ ${\rm cm^{-3}}$ and temperature $T = 10$ ${\rm K}$
\citep{Bonnor, Ebert1, Ebert2}, molecular clouds are not collapsing. No
large-scale velocities characteristic of collapse (or expansion) are observed. Taking into
account the 
contribution of the magnetic field, Mouschovias (1976a,b) calculated equilibria
of initially uniform, cold clouds embedded in a constant-pressure
external medium and threaded by a frozen-in magnetic field. These are
oblate structures 
relative to the field lines that acquire an hour-glass shape. Mouschovias 
and Spitzer (1976) used these
equilibrium states to determine analytically the critical mass-to-flux ratio
above which the 
magnetic field cannot support a cloud against its self-gravity. Under typical
molecular cloud conditions ($B = 30$ ${\rm \mu G}$ and $n = 
10^3$ ${\rm cm^{-3}}$), their expression yields a critical mass for
collapse $ \approx 5 \times 10^{2}$ ${\rm M_{\odot}}$, consistent with
observations. That critical mass-to-flux ratio is $20 \%$ smaller than the value
calculated later by Nakano \& Nakamura (1978) for an infinitessimally thin, uniform 
disk of infinite radius.

Early on, supersonic turbulence was proposed either as a means of supporting
the clouds or explaining the observed linewidths. But this scenario has the
unattractive quality of high energy requirements to sustain the
motion, because such motions decay very rapidly \cite{Field70}. 
Since the observed linewidths are supersonic but subAlfvenic, they
were attributed to large-scale, non-linear, oscillations, which are
indistinguishable from long-wavelength, long-lived hydrodynamic waves \cite{TM75}.
Such waves, because of their slow dissipation, have low energy requirements
\cite{Arons}. Analyses of observed
linewidths show a power-law relation 
between the velocity dispersion and the size of the observed objects (Larson
1981; Leung, Kutner \& Mead 1982; Myers 1983), which was thought to be the
signature of Kolmogorov turbulence. But this relation was explained
quantitatively as a
consequence of virialized linewidths of hydromagnetic waves in magnetically
supported, self-gravitating clouds (Mouschovias 1987a; Mouschovias \& Psaltis
1995).  

The condition that the mass-to-flux ratio must exceed a critical
value for a cloud to collapse is
necessary but not sufficient. Other factors, such as external
pressure and internal turbulent pressure, may aid or prevent the collapse,
respectively. In nature, the mass-to-flux ratio of a molecular cloud as a whole
is not expected to be very different from the critical value. If the ratio were
much greater, the entire cloud would collapse very quickly. Velocities
characteristic of this kind of collapse are not observed and the
overall efficiency of star formation is generally low ($M_{\rm star}/M_{\rm
  cloud} \lesssim 0.1$) \citep{Frerking, Zhou, Andre}. If the mass-to-flux
ratio of a cloud as a whole were much smaller than the critical value, then
the star formation timescale would be too long and clouds would rarely, if ever 
form stars. Molecular clouds are typically observed to be slightly
subcritical \citep{Carr87,Loren89,Crut93,Heiles93,Crut96,Shu99}.\footnote{Crutcher
  (1999), after compiling measurements from a number of molecular
  clouds, concluded that the typical cloud mass-to-flux ratio is a
factor of two grater than the critical value. However, both he and Shu \etal (1999) 
recognize that the actual
magnetic field strength must on average be a factor of two greater
than the measured value, since only the line of sight component is
measured by the Zeeman effect. In addition, the average column density
measured is by a factor of two greater than the actual relevant value (perpendicular to 
a thin disk) due to
geometric effects. Applying these corrections, Shu \etal (1999) concluded
that the clouds are typically subcritical.}     

In the interiors of magnetically supported, subcritical clouds, stars can still
form when self-gravity drives neutral matter through magnetic field lines (and the
plasma) to form fragments (or cores), whose mass-to-flux ratio eventually
exceeds the critical value (Mouschovias 1977, 1978, 1979b). This process is
referred to as ambipolar diffusion. Ambipolar diffusion was first proposed by
Mestel and Spitzer (1956) as a means by which an interstellar cloud as a whole
would reduce its magnetic flux. However, Mouschovias (1978, 1979b) pointed out that {\em the
essence of ambipolar diffusion is a redistribution of mass in the central flux
tubes of a molecular cloud, not a loss of magnetic flux by a cloud
as a whole}. The ambipolar-diffusion timescale was shown to be
three orders of magnitude smaller in the interior of a cloud than in the
outermost envelope, where the degree of ionization is much greater, leading to a
better coupling between neutral particles and the magnetic field (Mouschovias 1979b;
1987a,b). This
suggests a self-initiated fragmentation (or core formation) on the
ambipolar-diffusion timescale, which is less than $2\times 10^6$ yr at a density of 
$10^4$ ${\rm cm^{-3}}$ (see Mouschovias 1987a, eq. [12b]; or Mouschovias 1996, eq. [1]):
\[
\tau_{\rm AD} \approx \frac{\tau_{\rm ff}^2}{\tau_{\rm ni}} \approx
2 \times 10^6 \left( \frac{n_{\rm i}/n_{\rm H_2}}{10^{-7}} \right)  \ \ \  yr,
\]
where $\tau_{\rm ff}$ and $\tau_{\rm ni}$ are the free-fall and neutral-ion collision 
timescales, respectively. This expression is essentially independent of geometry 
(see Mouschovias 1987a). The notion aspoused by proponents of turbulence-driven star 
formation, namely, that ``rapid star formation'' is inconsistent with ambipolar-diffusion,
self-initiated star formation, has no merit. The issue is discussed quantitatively in 
another publication (Tassis \& Mouschovias 2004c).
The small degree of
ionization in cloud interiors also allows ambipolar diffusion to dump the
hydromagnetic waves or magnetic turbulence (thus rendering them irrelevant to 
star formation) over a
characteristic lengthscale comparable to the critical thermal lengthscale
($\approx$ Jean's length), thereby also leading to the thermalization of linewidths
in agreement with observations by Myers \& Benson (1983) and Myers, Linke \& Benson
(1983). This point of view is substantiated by
multifluid MHD numerical calculations in rectilinear (Paleologou \&
Mouschovias 1983; Mouschovias, Paleologou \& Fiedler 1985), cylindrical
(Mouschovias \& Morton 1991, 1992a,b) and axisymetric geometries (Fiedler \&
Mouschovias 1992, 1993; Ciolek \& Mouschovias 1993; Morton, Mouschovias \&
Ciolek 1994; Ciolek \& Mouschovias 1994; Desch \& Mouschovias 2001; Eng \&
Mouschovias 2004).

For a cloud as a whole and for a core during its
ambipolar-diffusion-controled, quastistatic (or subcritical) phase of contraction, magnetic
braking is very effective and keeps the core and the cloud essentially
corotating with the background (Mouschovias 1977, 1979a;
Mouschovias \& Paleologou 1979, 1980;
Mouschovias \& Morton 1985a,b; Basu \&
Mouschovias 1994, 1995a,b). Hence the centrifugal forces have a negligible
effect on the evolution of the contracting core. Magnetic braking is
responsible for allowing clouds to form in the first place and for reducing 
the angular momenta of cloud cores to their observed
low values, as well as for resolving the angular momentum problem of star
formation during the early, isothermal stage of contraction, as required by observations 
(see review by Mouschovias 1981, \S 6) 

After the central mass-to-flux ratio exceeds the critical value, the now
thermally and magnetically supercritical core begins to contract dynamically,
while the envelope remains magnetically supported. The first axisymmetric
calculations to follow both the 
quasistatic and the dynamic phase of core formation and contraction (Fiedler
\& Mouschovias 1992, 1993) found that force balance along field lines
is established rapidly and 
maintained even while dynamical contraction takes place perpendicular to the
field lines. A disklike geometry is established early and maintained
throughout the evolution of the protostellar fragment.
This result was
exploited by Ciolek \& Mouschovias (1993, 1994) and Basu \& Mouschovias
(1994, 1995a,b), who assumed force balance along field lines and integrated the
equations analytically over the vertical structure of the disk, thus reducing
the dimensionality of the problem for computational purposes without losing
any of the essential physics of the problem. 

These axisymmetric simulations have shown that, even during the dynamical 
phase of contraction, the infall of a core does not evolve into free fall
(see review by Mouschovias 1995). Even at high central densities ($n=10^{10}$
${\rm cm^{-3}}$), the maximum infall acceleration does not exceed $30\%$ that
of gravity, and the magnetic force dominates the thermal-pressure force
almost everywhere in the supercritical core except in a small central region. 
During the dynamical
contraction phase, both magnetic flux and angular momentum tend to get trapped
inside the core, with a consequent increase of the magnetic field and angular
velocity with gas density as $B_c \propto \rho_c^{1/2}$ and $\Omega_c \propto
\rho_c^{1/2}$ (Mouschovias 1976b, 1979a, 1989; Fiedler \& Mouschovias 1993;
Basu \& Mouschovias 1994). 

Millimeter and sub-millimeter
continuum observations \citep{WT94, Andre96, Bacmann2000} yield density
profiles for starless cores that are in agreement with those predicted
by the ambipolar diffusion models. Specific models
constructed for the Barnard 1 cloud (Crutcher \etal 1994) and L1544
(Ciolek \& Basu 2000) found the core properties predicted by the
ambipolar diffusion model to be in excellent agreement with the
observed values. In addition, Benson, Caselli, \& Myers (1998) measured
ion-neutral drift speeds consistent with the theoretical predictions. 

The leftover magnetic flux at the end of the early, isothermal stages of
contraction ($n \lesssim 10^{10}$ ${\rm cm^{-3}}$) exceeds observed
stellar magnetic fluxes and must therefore be either dissipated or somehow excluded from the matter destined to give birth to stars.
Approximate theoretical estimates argued that
Ohmic dissipation is effective in destroying the magnetic flux at
densities above $10^{10}$ ${\rm cm^{-3}}$ (Pneuman \& Mitchell 1965; Nishi, Nakano \&
Umebayashi 1991). Recent dynamical calculations by Desch and Mouschovias
(2001), which follow the evolution up to densities $n = 2 \times 10^{12}$ ${\rm
  cm^{-3}}$, show that ambipolar diffusion is much more effective than Ohmic
dissipation in decoupling the magnetic field from the neutral
matter. (Earlier work overestimated the ${\rm e-H_2}$ elastic
collision cross section, and hence the ohmic resistivity, by a factor
of 100; see review by Mouschovias 1996).

The evolution timescale of the dynamically
contracting supercritical core at the end of the above simulations is
less than 100 yr; a protostar
in hydrostatic equilibrium is expected to form soon thereafter. There have been several studies of the
accretion phase following the formation of a protostellar
core. Tomisaka (1996) performed a two-dimensional, axisymetric, isothermal
simulation and used the ``sink cell method'' to study the accretion
phase. He started from a supercritical initial condition and
assumed flux-freezing throughout the simulation.
Shu and collaborators extended the singular isothermal sphere
similarity solution \cite{Shu77}. Galli \& Shu (1993a,b) included
a spatially uniform, dynamically weak magnetic field, while Li \& Shu
(1997) studied the accretion of a singular isothermal disk with a
uniform mass-to-flux ratio, assuming the flux to be frozen in the
neutral fluid. In addition, in these studies the authors always assume
that the accretion onto the protostellar core starts from a static
singular configuration. This assumption is inconsistent both with the
aforementioned analytical and numerical studies of the evolution before
the formation of a hydrostatic protostellar core, which find
supercritical cores collapsing dynamically, and with
observations that find significant infall velocities in prestellar
cores \citep{tafalla98,W99}.
Safier, McKee \& Stahler (1997) used a spherical model to study
the collapse of both a prestellar core as well as the accretion phase
after the formation of a protostellar core. They neglected the thermal
pressure and the magnetic tension
force and they included ambipolar
diffusion in a phenomenological way without solving the induction
equation for the self-consistent evolution of the magnetic field.
Their spherical model was extended by Li (1998a, 1998b) to include the
thermal pressure and the induction equation. However, these models do
not satisfy the solenoidal condition
on the magnetic field and cannot account for the magnetic tension force.
 
Taking into account the decoupling of
the magnetic field from the matter Li \& McKee (1996) suggested that
Ohmic dissipation would halt the accretion of flux onto the protostar
and give rise to a hydromagnetic shock. However, Ciolek \& K\"{o}nigl (1998), 
extending the previous numerical studies of Mouschovias and
collaborators to study the later accretion phase of star formation by
using the ``central sink cell'' approach and assuming the magnetic
flux to be frozen in the ion fluid, found that ambipolar
diffusion is sufficiently effective at much lower densities (or much
larger radii) to halt flux advection and drive a hydromagnetic
shock, independent of the effect of Ohmic dissipation. This
hydromagnetic shock is also present in the self-similar solution
obtained by Contopoulos, Ciolek \& K\"{o}nigl (1998). Starting from a
dynamical initial state (nonzero infall velocity) and assuming force
balance in the $z$-direction as well as a power-law scaling of the
density of ions with respect to the density of neutrals, they were able
to obtain a similarity solution which is very close to the numerical
results of Ciolek \& K\"{o}nigl (1998). 

In this paper we formulate the problem of the late accretion phase of an 
isothermal magnetic disk surrounding a protostar by taking into 
account important physics ignored by
previous calculations: (1) the decoupling of the 
ions from the magnetic field lines, which occurs at densities above 
$10^8 \, {\rm cm ^{-3}}$ (Desch \& Mouschovias 2001) and which leaves 
the magnetic field frozen in the much less massive and much more
tenuous electron fluid alone; (2) the chemical and dynamical effects
of the positively-charged grains, 
the abundance of which becomes significant in the same density regime. 
The results of this calculation are discussed in a companion paper
(Tassis \& Mouschovias 2004; hereafter Paper II).

In \S 2 we describe the model cloud. In \S 3 we discuss the six-fluid
MHD equations appropriate for the physical system under study, and in 
\S 4 the thin-disk approximation employed in our numerical code. In \S 5 
we state our initial and boundary conditions, and in \S 6 we discuss
in detail the implementation of the central sink approximation. The 
dimensionless equations solved numerically and the method of 
solution are described in \S 7. We conclude in \S 8 by discussing 
the most important differences of our formulation from previous work
applied to similar problems.

\section{MODEL CLOUD}

We consider a self-gravitating, magnetic, weakly-ionized model molecular cloud
consisting of neutral particles (${\rm H_2}$ with $20\%$ ${\rm He}$
by number), ions (both molecular ${\rm HCO^+}$ and atomic ${\rm
  Na^+}$, ${\rm Mg^+}$), electrons, neutral grains, singly negatively-charged
and positively-charged grains. Following Desch and Mouschovias (2001),
the abundances of charged species is determined from the chemical reaction
network shown in Table \ref{table1}. (For a detailed discussion of the chemistry,
see Ciolek \& Mouschovias 1993, 1995). The positively-charged grains
cannot be neglected at densities grater than $10^9 \,
{\rm cm^{-3}}$, since their abundance becomes comparable to
that of negative grains (see Fig. 2 in paper II).
For simplicity, we consider spherical grains of uniform size. Desch
and Mouschovias (2001) used a chemical model that accounted for grains of
different sizes and found that the effect of the grain size distribution on
the evolution of the cloud is minimal. In the
case of collisions of ions (molecular or atomic) with grains, we assume that
the ions do not get attached to the grains, but that they get
neutralized, with the resulting neutral particle escaping into the gas
phase. Thus the total abundance of metals as well as the ${\rm
  HCO}$ abundance remain constant. Grain growth and depletion of elements by
attachment onto grains are not considered in this paper.

The axisymmetric model cloud is initially
in an exact equilibrium state with self-gravity and external pressure
being balanced by internal magnetic and thermal-pressure forces. 
We follow the ambipolar-diffusion--initiated evolution up to a central
density of $10^{11}$
${\rm cm^{-3}}$, below which isothermality is a good approximation. 
To study the later stages of the evolution of the disk surrounding the
forming star, we employ the ``central sink cell'' method, originally
used by Boss \& Black (1982) and more recently employed by Ciolek \&
K\"{o}nigl (1998). The radius of the
central sink defines the inner boundary
$r=R_{\rm inner}$ of the computational region. The use of a central
sink makes the study of the dynamics of the collapsing
protostellar fragment possible, even after the formation of an opaque core,
by fixing the value of $R_{\rm inner}$ at a radius beyond which the density is low enough
($n_{\rm n} \leq 10^{11}$ ${\rm cm^{-3}}$) that isothermality remains
an excellent approximation. Computationally taxing radiative transfer becomes unnecessary.

\section{SIX-FLUID MHD DESCRIPTION OF MAGNETIC STAR FORMATION}

We extend the multifluid formalism described by Ciolek and Mouschovias
(1993) to include the positively charged grains and to relax the assumption of 
flux-freezing in the ions. Only the electrons are assumed to be attached to the 
magnetic field lines. The equations governing the
behavior of the six-fluid system (neutrals, electrons, ions,
negative, positive, and neutral grains) are:

\begin{mathletters}
\beq
\frac{\partial \rho_{\rm n}}{\partial t} + \nabla \cdot (\rho_{\rm n}
\bvec{v}_{\rm n})=0 ,\label{firsteq}
\eeq
\beq
\frac{\partial (\rho_{\rm g_0} + \rho_{\rm g_-} +\rho_{\rm g_+})}{\partial t} + \nabla
\cdot  (\rho_{\rm g_0} \bvec{v}_{\rm g_0} + \rho_{\rm g_-} \bvec{v}_{\rm g_-}
+ \rho_{\rm g_+}
\bvec{v}_{\rm g_+})=0 ,\label{gcont}
\eeq
\beq
\frac{\partial ( \rho_{\rm n} \bvec{v_{\rm n}}) }{\partial t} +
\nabla \cdot (
\rho_{\rm n}  \bvec{v_{\rm n}}  \bvec{v_{\rm n}}) = \rho_{\rm n} \bvec{g} - \nabla
P_{\rm n} + \frac{1}{4\pi} (\nabla \times \bvec{B}) \times
\bvec{B},\label{nforceq}
\eeq
\beq
0 =  -en_{\rm e}( \bvec{E} + \frac{\bvec{v_{\rm e}}}{c} \times \bvec{B}) +
\bvec{F}_{\rm en}, \label{eforceq}
\eeq
\beq
0  =  {\rm e} n_{\rm i}( \bvec{E} + \frac{ \bvec{v_{\rm i}}}{c}  \times \bvec{B}) +
\bvec{F}_{\rm in}, \label{iforceq}
\eeq
\beq
0  =  -{\rm e} n_{\rm g_{-}}( \bvec{E} + \frac{ \bvec{v_{\rm g_{-}}}}{c} \times \bvec{B}) +
\bvec{F}_{\rm g_{-}n} + \bvec{F}_{\rm g_{-}g_{0},inel},
\label{gmforceq}
\eeq
\beq
0  =  en_{\rm g_{+}}( \bvec{E} + \frac{ \bvec{v_{\rm g_{+}}}}{c} \times \bvec{B})
+\bvec{F}_{\rm g_{+}n} + \bvec{F}_{\rm g_{+}g_{0},inel}, 
\label{gpforceq}
\eeq
\beq
0  =  \bvec{F}_{\rm g_{0}n} + \bvec{F}_{\rm g_{0}g_{-},inel}+
\bvec{F}_{\rm g_{0}g_{+},inel},\label{g0forceq}
\eeq
\beq
\nabla \times \bvec{B} =  \frac{4\pi}{c} \bvec{j},\label{amp}
\eeq
\beq
\bvec{j} =  e( n_{\rm i} \bvec{v}_{\rm i} + n_{\rm g_+} \bvec{v}_{\rm
  g_+} - n_{\rm e} \bvec{v}_{\rm e} - n_{\rm g_-} \bvec{v}_{\rm g_-}
),\label{jdef}
\eeq
\beq
\frac{\partial \bvec{B}}{\partial t}  =  - c (\nabla \times
\bvec{E}),\label{faradayA}
\eeq
\beq
\nabla \cdot \bvec{g}  =  - 4 \pi G \rho_{\rm n},
\eeq
\beq
P_{\rm n}  =  \rho_{\rm n} C_{\rm n}^2.\label{lasteq}
\eeq
\end{mathletters}
The quantities $\rho_{\rm s}$, $n_{\rm s}$ and $\bvec{v}_{\rm s}$ denote the mass
density, number density, and velocity of species {\it s}. The subscripts
n, i, e, $\rm g_{-}$, $\rm g_{+}$ and $\rm g_{0}$ refer to the neutrals, ions,
electrons, negatively-charged grains, positively-charged grains, and neutral
grains, respectively. The symbols $\bvec{g}$, $\bvec{E}$ and $\bvec{B}$ denote
the gravitational, electric and magnetic fields, respectively.
The magnetic field satisfies
the condition $\nabla \cdot \bvec{B} = 0$ everywhere at all
times.

The frictional force (per unit volume) on species $s$ due to elastic
collisions with neutrals is given by
\beq
\bvec{F}_{s {\rm n}} = \frac{\rho_{s}}{\tau_{s
    n}}(\bvec{v}_{n}-\bvec{v}_{s}), \ \ \ \ \ \ \ \ \ \ \ s = {\rm i}, \ {\rm e},
\ {\rm g_-}, \ {\rm g_+}, \ {\rm g_0},  
\eeq
where the mean collision time, accounting for both $s$ - ${\rm H_{\rm 2}}$ and $s$
- He collisions, is
\beq
\tau_{s{\rm n}} = \frac{\tau_{s {\rm H_2}}}{a_{s{\rm He}}} =
\frac{1}{a_{s{\rm He}}} \frac{m_{\rm H_2} + m_{s}}{\rho_{\rm H_2} \langle \sigma
  w \rangle_{s{\rm H_2}}} , \ \ \ \ \ \ \ \ \ \ s = {\rm i}, {\rm e},
{\rm g_-}, {\rm g_+},{\rm  g_{\rm 0}},
\eeq
and the quantity
\beqar
a_{s{\rm He}} & = & 1.14 \ \ \ \ \ \ {\rm for} \ \ s = {\rm i}, \\
              & = & 1.16 \ \ \ \ \ \ {\rm for} \ \ s = {\rm e}, \\
              & = & 1.28 \ \ \ \ \ \ {\rm for} \ \ s = {\rm g_-, g_+} \ {\rm or} \ {\rm g_0} 
\eeqar
is the factor by which the presence of He reduces the slowing-down time of
species $s$ with respect to $s$ - ${\rm H_2}$ collisions alone (see
Mouschovias 1996, \S 2.1). The mean collisional rate between ions of
mass $m_{\rm i}$ and hydrogen molecules of mass $m_{\rm H_2}$ 
is $\langle \sigma w \rangle_{\rm i H_2} = 1.69\times10^{-9}$
 ${\rm cm^{3}}$ ${\rm s^{-1}}$ for ${\rm HCO^+ - H_2}$ collisions, and
almost identical to this value for
 ${\rm Na^+ - H_2}$ and ${\rm Mg^+ - H_2}$ collisions. The value of the mean
 collisional rate between 
electrons and ${\rm H_2}$ is $\langle \sigma w \rangle_{\rm e H_2} = 1.3 \times 10^{-9}$
${\rm cm^{3}}$ ${\rm s^{-1}}$. In calculating these collisional rates, the Langevin
approximation is used for ion-neutral collisions but not for electron-neutral 
collisions, for which the electron spin is important \cite{m96}. The grain-${\rm H_2}$
collisional rate is given by 
\beq
\langle \sigma w \rangle_{\rm g H_2} = \pi a^2 (8k_{\rm B}T/\pi m_{\rm H_2})^{1/2}.
\eeq 
This expression is valid only if the velocity difference between a grain and a
hydrogen molecule is smaller than the sound speed in the neutrals
\cite{Ciolek93}. Otherwise, one has to use
\beq
\langle \sigma w \rangle_{\rm g H_2} = \pi a^2 |v_{\rm n} - v_{\rm g}|.
\eeq

The quantity $\bvec{F}_{\gamma \delta,{\rm inel}(k)}$ in equations
(\ref{gmforceq}) - (\ref{g0forceq}) is the force (per unit
volume) on grain fluid $\gamma$ due to the 
conversion of dust particles of fluid $\delta$ into dust particles of
fluid $\gamma$, with mass gain $\delta m_{\gamma}$, (note that $m_{\gamma} = m_{\delta}$) 
via an inelastic process {\it k}, where {\it k} can be any one of the
processes listed in Table \ref{table1} that involve grains. The
momentum transferred to fluid $\gamma$  from fluid $\delta$ is 
\beq
\bvec{\Delta p} = |\delta m_{\gamma}| (\bvec{v}_{\delta} - \bvec{v}_{\gamma});
\eeq  
and we can write
\beq
\bvec{F}_{\gamma \delta,{\rm inel}(k)} = m_{\gamma} \left| \frac{\partial
  n_{\gamma}}{\partial t} \right|_{k} (\bvec{v}_{\delta} - \bvec{v}_{\gamma}),
\eeq 
where the time rate of change of the number density of $\gamma$ due to the
inelastic process ({\it k}) is given by
\beqar
\left. \frac{\partial n_{\gamma}}{\partial t} \right| _{({\it k})} & = &
n_{\eta} n_{\delta} \langle \sigma w \rangle_{({\it k})} \nonumber \\
 & = &  \frac{n_{\delta}}{\tau_{\delta \eta , {\rm inel}}},
\eeqar 
and $\tau_{\delta \eta , {\rm inel}}$ is the timescale for a particle $\delta$ to
find a particle $\eta$ and be converted into $\gamma$.
Thus, in general, since $m_{\gamma} = m_{\delta}$
\beq
\bvec{F}_{\gamma \delta,{\rm inel}({\it k})} =  \frac{
  \rho_{\delta}}{ \tau_{\delta \eta , {\rm inel}}} (\bvec{v}_{\delta} - \bvec{v}_{\gamma}).
\eeq 
Furthermore, according to Newton's third law, the fluid $\delta$ will also
experience an equal and opposite force 
\beq
\bvec{F}_{\delta \gamma,{\rm inel}({\it k})} = - \bvec{F}_{\gamma
  \delta,{\rm inel}({\it k})} = \frac{
  \rho_{\delta}}{ \tau_{\delta \eta , {\rm inel}}} (\bvec{v}_{\gamma} - \bvec{v}_{\delta}).
\eeq 
The relevant timescales due to these inelastic processes are
$\tau_{\rm g0 e , {\rm inel}} = (n_{\rm e} \alpha_{\rm eg0})^{-1}$, $\tau_{\rm g0 i ,
  {\rm inel}} = (n_{\rm i} \alpha_{\rm ig0})^{-1}$, $\tau_{\rm g_- i ,
  {\rm inel}} = 
(n_{\rm i} \alpha_{\rm ig_-})^{-1}$,  $\tau_{\rm g_+ e ,
{\rm  inel}} = (n_{\rm e} \alpha_{\rm eg_+})^{-1}$,  $\tau_{\rm g_-
  g_+ , {\rm inel}} =
(n_{\rm g_+} \alpha_{\rm g_- g_+})^{-1}$,  $\tau_{\rm g_+ g_- , {\rm inel}} = (n_{\rm g_-}
\alpha_{\rm ig_-})^{-1}$, 
where the reaction rates $\alpha_{\rm eg0}$, $\alpha_{\rm ig0}$, $\alpha_{\rm ig_-}$,
$\alpha_{\rm eg_+}$ and $\alpha_{\rm g_- g_+}$ are given in Appendix A.

In the force equations of the electrons, ions, and grains, the
acceleration terms have been neglected because of the small inertia of these
species. Mouschovias, Paleologou \& Fiedler (1985) included the
acceleration term for the plasma and showed that the plasma reaches a terminal drift
velocity very fast.
Likewise, the thermal-pressure and gravitational forces (per unit volume) have
been dropped from the force equations of all species other than the neutrals
because they are negligible compared to the electromagnetic and collisional
forces. The inelastic momentum losses by the electron and ion fluids due to
attachment onto grains and neutralization are negligible compared
to the momentum loss due to elastic collisions, and they have been omitted from
the force equations (\ref{eforceq}) and (\ref{iforceq}). 

Finally, since we constrain our study to the region characterized by
densities smaller than $10^{11}$ ${\rm cm^{-3}}$, 
isothermality is a good approximation, and 
$C_{\rm n} = (k_{\rm B} T/ \mu m_{\rm H})^{1/2}$ is the isothermal speed of sound in the
neutrals, $k_{\rm B}$ the Boltzmann constant, and $\mu$ the mean mass per
neutral particle in units of the atomic-hydrogen mass, $m_{\rm H}$.
For the early stages of star formation or for the
isothermally infalling gas outside the optically thick inner region, the
temperatures are much smaller than the
grain sublimation temperature $\approx 1500$ K, and a mass continuity
equation (\ref{gcont}) for the grains (charged + neutral) is appropriate.

Altogether, then, we have a system of 13 equations, (\ref{firsteq})-(\ref{lasteq}), which contain
17 unknowns ($\rho_{\rm n}$, $P$, $\bvec{E}$, $\bvec{B}$, $\bvec{j}$, $\bvec{g}$,
$\bvec{v}_{\rm n}$, $\bvec{v}_{\rm e}$, $\bvec{v}_{\rm i}$, $\bvec{v}_{\rm
  g_-}$, $\bvec{v}_{\rm g_+}$,
$\bvec{v}_{\rm g_0}$, $\rho_{\rm e}$, $\rho_{\rm i}$, $\rho_{\rm g_-}$,
$\rho_{\rm g_+}$,
$\rho_{\rm g_0}$). To close the system, the densities of electrons, ions and charged
grains ($n_{\rm e}$, $n_{\rm i}$, $n_{\rm g-}$, and $n_{\rm g+}$) are
calculated from the equilibrium chemical model described in
Appendix A.

Solving the electron force equation for the electric field, we obtain the
generalized Ohm's law
\beq\label{ohmA}
\bvec{E} = - \frac{\bvec{v}_{\rm e}}{c} \times \bvec{B} + \frac{\rho_{\rm
    e}}{\tau_{\rm en}}
\frac{(\bvec{v}_{\rm e} - \bvec{v}_{\rm n})}{e n_{\rm e}}.
\eeq
The second term on the right-hand side can be neglected for densities
$< 10^{14}$ ${\rm cm^{-3}}$, reducing equation (\ref{ohmA}) to the simple form
\beq\label{fluxfreezing}
\bvec{E} = - \frac{\bvec{v}_{\rm e}}{c} \times \bvec{B}.
\eeq
This is inserted in Faraday's law of induction (eq. [\ref{faradayA}]) to obtain the
flux-freezing equation in the electrons
\beq\label{fluxfrz}
\frac{\partial \bvec{B}}{\partial t} = \nabla \times(\bvec{v}_{\rm e} \times \bvec{B}). 
\eeq

\section{GOVERNING EQUATIONS IN THE THIN-DISK APPROXIMATION}

Numerical simulations of the ambipolar-diffusion induced evolution of
axisymmetric, isothermal, molecular clouds have shown that force
balance 
is rapidly established along field lines and is maintained throughout the evolution
of the model cloud, even during the dynamic phase of contraction
that follows the formation of a thermally and magnetically 
supercritical core (Fiedler \& Mouschovias
1992,1993; Desch \& Mouschovias 2001). Following the approach of Ciolek \& Mouschovias (1993),
we take advantage of these results and model the cloud as an axisymmetric thin
(but not infinitesimally thin) disk, with its axis of symmetry aligned
with the $z$-axis of a cylindrical polar
coordinate system $(r,\phi,z)$ and with force balance along the field lines
maintained at all times. The sense in which the thin-disk approximation is
used is that the radial variation of any physical quantity is small over a
distance comparable to the local disk half-thickness.
The upper and lower surfaces of the disk are at $z = +Z(r,t)$ and $z =
-Z(r,t)$, while the radius $R$ of the disk satisfies the condition $R \gg Z$. 
The magnetic field has the form
\begin{mathletters}
\beq
\bvec{B}(r,z,t) = \hat{z} B_{z,{\rm eq}}(r,t) \ \ \ \ \  \ \ \ \ \ \ \
\ \ \ \ \ \ \ \ \ \ \ \ {\rm for} \ \ \left|z
\right| \leq Z(r,t),
\eeq
\beq
\bvec{B}(r,z,t) = \hat{z} B_{z}(r,z,t) + \hat{r} B_{r}(r,z,t) \ \ \ \
\ \ {\rm for} \ \ \left|z
\right| > Z(r,t).
\eeq
\end{mathletters}
Inside the disk the magnetic field has only a $z$-component, equal to the field
in the equatorial plane $B_{z,{\rm eq}}$. The $r$-component of the field is antisymmetric
about the equatorial plane, i.e., $B_{r}(r,z) = -B_{r}(r,-z)$. Far away from the
cloud the magnetic field is assumed to be uniform, perpendicular to the plane
of the disk and constant in time. The normal component of the magnetic
field is continuous across the upper and lower surfaces of the disk. The
unit vector normal to the upper and lower surfaces has both $z$ and $r$ components
(i.e., the surface is not parallel to the equatorial plane). The model cloud is
embedded in an external medium of constant thermal pressure $P_{\rm ext}$.

For computational simplicity, we reduce the mathematical
dimentionality of the problem by integrating the equations
analytically over $z$, assuming $z$-independence
of all physical quantities. This is an excellent approximation since
thermal-pressure forces smooth out density gradients over lengthscales smaller than
the critical thermal lengthscale (see Mouschovias 1991), which is comparable to $Z(r,t)$. The thin-disk
equations governing the model cloud are
\begin{mathletters}
\beq\label{thin1}
\sigma_{\rm n}(r,t) = \int_{-Z(r,t)}^{+Z(r,t)} \rho_{\rm n} dz  =  2
\rho_{\rm n}(r,t)Z(r,t), 
\eeq
\beq\label{thin2}
\frac{\partial \sigma_{\rm n}}{\partial t} + \frac{1}{r}\frac{\partial (r 
\sigma_n v_{{\rm n},r})}{\partial r}  =  0, 
\eeq  
\beq\label{thin3}
\frac{\partial (\chi_{\rm g} \sigma_{\rm n})}{\partial t} + \frac{1}{r}\frac{\partial [r 
\sigma_{\rm n} (\chi_{\rm g-} v_{{\rm g-},r} + \chi_{\rm g+} v_{{\rm g+},r} +
\chi_{\rm g0}
v_{{\rm g0},r})]}{\partial r}  =  0, 
\eeq
\beq\label{thin4}
\rho_{\rm n}(r,t)C^2 = P_{\rm ext} + \frac{\pi}{2}G \sigma_{\rm n}^2(r,t),
\eeq
\beq\label{thin5}
\frac{\partial (\sigma_{\rm n} v_{{\rm n},r})}{\partial t} + \frac{1}{r}\frac{\partial( r 
\sigma_{\rm n} v_{{\rm n},r}^2)}{\partial r}  =  -C_{\rm eff}^2 \frac{\partial
\sigma_{\rm n}}{\partial r} + \sigma_{\rm n} g_r + F_{{\rm mag},r} ,
\eeq
\beq
C_{\rm eff}^2 = \frac{\pi}{2}G \sigma_{\rm n}^2 \frac{[3P_{\rm ext} + (\pi/2)G
  \sigma_{\rm n}^2]}{[P_{\rm ext} + (\pi/2)G\sigma_{\rm n}^2]^2} C^2,
\eeq
\beq
F_{{\rm mag},r} = \frac{\pi}{2} \left\{ B_{z,{\rm eq}}\left( B_{r,Z}-Z\frac{\partial
  B_{z,{\rm eq}}}{\partial r}\right) + \frac{1}{2} \frac{\partial Z}{\partial r}\left[B_{r,Z}^2
+ 2B_{z,{\rm eq}}\left(B_{r,Z}\frac{\partial Z}{\partial r}\right)+\left(B_{r,Z}\frac{\partial
  Z}{\partial r}\right)^2\right] \right\}, 
\eeq
\beq\label{thin6}
g_{\rm r}(r) = 2\pi G \int_{0}^{\infty}dr' r' \sigma_{\rm n}(r')\mathcal{M}(r,r'),
\eeq
\[
\mathcal{M}(r,r') = \frac{d}{dr}\left[ \int_{0}^{\infty} dk J_0(kr)J_0(kr')
\right]=\frac{2}{\pi}\frac{d}{dr}\left[ \frac{1}{r_>}K\left( \frac{r_<}{r_>}\right) \right],
\]
\beq\label{thin7}
B_{r,Z}(r) = - \int_{0}^{\infty}dr' r' [B_{z,{\rm eq}}(r')-B_{\rm ref}]\mathcal{M}(r,r'),
\eeq
\beq\label{thin8}
\frac{\partial B_{z,{\rm eq}}}{\partial t} + \frac{1}{r}\frac{\partial (r 
B_{z,{\rm eq}} v_{{\rm e},r})}{\partial r}  =  0, 
\eeq
\beq\label{theevel}
v_{\rm e} = v_{\rm n} + \frac{F_{\rm mag}\tne}{\sigma_{\rm n}\Lambda_{\rm e}}
\eeq
%%%%%%%%%%%%%%%%%%%%%%%%%%%%%%%%%%%%%%%%%%%%%%%%%%%%%%%%%%%%%%%%%%%%%%%%%%%%%%%%%%%%%%%

\hfill\begin{tabular}{ll}\label{thevels}
$\displaystyle \vi = \frac{1}{\Theta_{\rm i}+1}\vn + 
\frac{\Theta_{\rm i}}{\Theta_{\rm i}+1}\ve\label{soli}$,&
$\displaystyle \vif =  \Xi_{{\rm i},\phi}\left(\frac{1}{\Theta_{{\rm i},\phi}+1}\vn + 
\frac{\Theta_{{\rm i},\phi}}{\Theta_{{\rm i},\phi}+1}\ve\right) \label{solif}$,
\end{tabular}\hfill(\stepcounter{equation}\theequation,m)
\stepcounter{equation}

\hfill\begin{tabular}{ll}
$\displaystyle \vgm = \frac{1}{\Theta_{\rm g-}+1}\vn + 
\frac{\Theta_{\rm g-}}{\Theta_{\rm g-}+1}\ve\label{solm}$,&
$\displaystyle \vgmf = \Xi_{{\rm g-},\phi}\left(\frac{1}{\Theta_{{\rm g-},\phi}+1}\vn + 
\frac{\Theta_{{\rm g-},\phi}}{\Theta_{{\rm g-},\phi}+1}\ve\right) \label{solmf}$,
\end{tabular}\hfill(\stepcounter{equation}\theequation,o)
\stepcounter{equation}

\hfill\begin{tabular}{ll}
$\displaystyle \vgp=  \frac{1}{\Theta_{\rm g+}+1}\vn + 
\frac{\Theta_{\rm g+}}{\Theta_{\rm g+}+1}\ve\label{solp}$,&
$\displaystyle \vgpf = \Xi_{{\rm g+},\phi}\left(\frac{1}{\Theta_{{\rm g+},\phi}+1}\vn + 
\frac{\Theta_{{\rm g+},\phi}}{\Theta_{{\rm g+},\phi}+1}\ve\right)\label{solpf}$,
\end{tabular}\hfill(\stepcounter{equation}\theequation,q)
\stepcounter{equation}

\hfill\begin{tabular}{ll}
$\displaystyle \vgo = \frac{1}{\Theta_{\rm g0}+1}\vn + 
\frac{\Theta_{\rm g0}}{\Theta_{\rm g0}+1}\ve\label{soln}$,&
$\displaystyle \vgof =  \Xi_{{\rm g0},\phi}\left(\frac{1}{\Theta_{{\rm g0},\phi}+1}\vn + 
\frac{\Theta_{{\rm g0},\phi}}{\Theta_{{\rm g0},\phi}+1}\ve\right)\label{solnf}$,
\end{tabular}\hfill(\stepcounter{equation}\theequation,s)
\stepcounter{equation}

\hfill\begin{tabular}{l}
$\displaystyle \vef =  \Xi_{{\rm e},\phi}\left(\frac{1}{\Theta_{{\rm e},\phi}+1}\vn + 
\frac{\Theta_{{\rm e},\phi}}{\Theta_{{\rm e},\phi}+1}\ve\right)\label{solef}$.
\end{tabular}\hfill(\stepcounter{equation}\theequation)
\end{mathletters}

\noindent
All symbols introduced in equations (\ref{theevel})-(\ref{thevels}t)
are defined in Appendix B.
A detailed derivation of the first nine thin-disk equations,
(\ref{thin1})-(\ref{thin7}), is
given in Ciolek \& Mouschovias (1993).
Introducing the effective sound speed, $C_{\rm eff}$, we account
for the contribution of the radial force exerted by the external pressure
$P_{\rm ext}$ on the upper and lower surfaces of the disk, since the latter
are not horizontal. In the expression for the total radial
magnetic force (per unit area), $F_{{\rm mag},r}$, includes the magnetic tension force (the
$B_r$ terms) acting on the surfaces of the disk.
The $r$-component of the gravitational field is calculated from Poisson's equation
for a thin disk (\ref{thin6}). $J_0$ is the zeroth-order Bessel function of
the first kind, $K$ is the complete elliptic integral of the first kind
\cite{AS}, $r_< = {\rm min}[r,r']$, and $r_> = {\rm max}[r,r']$.

Equations (\ref{thevels}l-t) give the
velocities of the ions and the three grain species in the $r$
and $\phi$ directions, and the $\phi$-component of the electron
velocity in terms of the radial velocities of the
neutrals ($v_{\rm n}$) and electrons ($v_{\rm e}$; eq. [\ref{theevel}]). They come from the solution of the
steady-state force equations of the species and Amp\`{e}re's law 
(eqs. [\ref{eforceq}]-[\ref{g0forceq}] and [\ref{amp}],
where $\bvec{j}$ in eq. [\ref{amp}] is given by eq. [\ref{jdef}]).
These are algebraic equations, but
they have coupled $r$ and $\phi$ components due to the presence of the Lorentz
terms (no $z$-components since we assume $z$-independence for all
quantities inside the model cloud). All the {\em radial} velocities,
except that of the electrons, have the form 
\beq
v_s = \frac{\Theta_s}{\Theta_s +1}\ve + \frac{1}{\Theta_s+1}\vn 
\eeq
(see Appendix A), where 
$\Theta _s$ is the {\it indirect attachment parameter} of species $s$ : for $\Theta_s \gg
1$ the $r$-velocity of species $s$ is approximately equal to that of the electrons, and
the species is well attached to the field lines, while for $\Theta_s \ll 1$
the species is detached and is falling in with the neutrals ($v_s
\approx \vn$).

\section{INITIAL AND BOUNDARY CONDITIONS}

The model cloud is initially in equilibrium, with magnetic and
thermal-pressure forces balancing the gravitational and external-pressure
forces, in the absence of ambipolar diffusion. Any
evolution at all is the result of the onset of ambipolar diffusion.
The values of the free parameters are obtained from an initial reference
state of the model cloud which is allowed to relax to an equilibrium
configuration in the absence of ambipolar diffusion.
The initial equilibrium state is thermally supercritical but magnetically
subcritical. A detailed description of the calculation of the initial
equilibrium state from a reference state is given in Ciolek \& Mouschovias
(1993). Magnetically critical or supercritical models have been
studied by Fiedler \& Mouschovias (1992) and Basu \& Mouschovias
(1995a, 1995b). 

At the outer boundary of the disk we require continuity of thermal as
well as magnetic pressures
\begin{mathletters}
\beq \label{out_bc1}
\rho_{\rm n}(R,t) = \frac{P_{\rm ext}}{C^2},
\eeq
and 
\beq \label{out_bc2}
B_{z}(R,t) = B_{z,{\rm ext}}(R,t).
\eeq
On the axis of symmetry, radial components of all velocities vanish

\hfill \begin{tabular}{ccc}
$
v_{\rm n}(r=0) = 0$, &
$v_{\rm e}(r=0) = 0$, &
$v_{\rm i}(r=0) = 0$,
\end{tabular} \hfill
(\stepcounter{equation}
\theequation,d,e)

\stepcounter{equation}
\stepcounter{equation} 
\hfill
\begin{tabular}{ccc}
$v_{\rm g-}(r=0) = 0$, &
$v_{\rm g+}(r=0) = 0$, &
$v_{\rm g0}(r=0) = 0$.
\end{tabular} \hfill
(\stepcounter{equation}
\theequation,g,h)\\
\stepcounter{equation}
\stepcounter{equation}
In addition, the radial components of all forces vanish on the axis, which
implies that

\hfill\begin{tabular}{cc}
$g_r(r=0) = 0$,&
$B_{r,Z}(r=0) = 0$,
\end{tabular}\hfill(\stepcounter{equation}\theequation,j)
\stepcounter{equation}

\noindent and

\hfill
\begin{tabular}{ccc}
$\displaystyle \left.\frac{\partial \sigma_{\rm n}}{\partial r} \right|_{r=0} = 0$,&
$\displaystyle \, \, \left.\frac{\partial B_{z,{\rm eq}}}{\partial r} \right|_{r=0}= 0$, 
$\displaystyle \, \, \, \, \left. \frac{\partial Z}{\partial r} \right|_{r=0} = 0$. \end{tabular}
\hfill(\stepcounter{equation}\theequation,l)
\end{mathletters}

\section{THE CENTRAL-SINK APPROXIMATION}

In order to follow the evolution up to the time that 1 $M_{\odot}$
accumulates in a central region only a few AU in radius, $R_{inner}$, some of the thin-disk equations must be modified
and new equations must be added. The rate at which mass accumulates in this central
region, is given by
\begin{mathletters}
\beq \label{csa1}
\frac{\partial M_{*}}{\partial t} = -2 \pi r \sigma_{\rm n}\left. v_{\rm
  n} \right|_{r=R_{\rm inner}}. 
\eeq
Similarly, the rate of accretion of grain mass is
\beq\label{csa2}
\frac{\partial M_{\rm g,*}}{\partial t} = -2 \pi r \left. (\sigma_{\rm g-} v_{\rm g-}+
\sigma_{\rm g+}v_{\rm g+} + \sigma_{\rm g0}v_{\rm g0})\right|_{r=R_{\rm inner}}, 
\eeq
and that of magnetic flux
\beq\label{csa3}
\frac{\partial \Phi_{\rm B,*}}{\partial t} = -2 \pi r
\left. B_{z,{\rm eq}}v_{\rm e} \right|_{r=R_{\rm inner}}, 
\eeq
\end{mathletters}
where $M_{\rm *}(t)$, $M_{\rm g,*}(t)$ and $\Phi_{\rm B,*}(t)$ are the
mass of gas, the mass of grains, and the magnetic flux in the central sink,
respectively, destined to become a star. These three equations must be solved simultaneously with the
other thin-disk equations. 

Also, the expressions for the radial component of the gravitational and magnetic fields
must be modified in order to incorporate the influence of the mass and
magnetic flux of the central cell, which, although not a part of the
active computational region, affects the matter at $r 
\ge R_{\rm inner}$ gravitationally and magnetically.
No information about the spatial variation of the column
density inside the central sink is available. However, we may account
for the formation of a dense core (at the center) at later times by expressing
the gravitational field as
\beq\label{grav_field}
g_r(r) = -\frac{GM_{cent}}{r^2} + 2\pi G \int_{0}^{\infty}dr' r'
\sigma_n(r')M(r,r').
\eeq
The first term on the RHS is the contribution of the mass in the central sink 
assuming that it is spherical in shape. In reality, the lower limit of the integral in
the second term on the RHS of equation (\ref{grav_field}) should be $R_{\rm
  inner}$. However, for computational convenience we extend the integral to
$r=0$, and set 
$\sigma_{\rm n}(r<R_{\rm inner})= \sigma_0$, where $\sigma_0 \ll
\sigma_{\rm n}(r=R_{\rm inner})$, so as to avoid counting the mass inside the central
sink twice.

The radial component of the magnetic field on the surface of the disk, which
appears in the magnetic tension force, is still obtained from the expression
\beq
B_{r,Z}(r) = - \int_{0}^{\infty}dr' r' [B_{z,{\rm eq}}(r')-B_{\rm ref}]M(r,r'),
\eeq
but $B_{z,\rm eq}(r < R_{\rm inner})$ is now calculated by assuming that the flux 
inside the central sink is uniformly distributed. (Ciolek \& K\"{o}nigl 1998 used eq. [23]
with the lower limit of the integral being $R_{\rm inner}$ and they added a monopole term 
on the right-hand side to mimic the effect of the magnetic flux in the central sink.)

The presence of a central point-like mass affects the equilibrium along the
magnetic flux tubes as well. The equation of force balance in the $z$-direction
now becomes
\beq\label{csazfb}
\rho_{\rm n}(r,t)C^2 = P_{\rm ext} + \frac{\pi}{2}G \sigma_{\rm n}^2(r,t) +
\frac{GM_{\rm cent}\rho_{\rm n}}{r} \left\{ 1 - \left[ 1 + \left(
        \frac{\sigma_{\rm n}}{2\rho_{\rm n} r} \right)^2 \right]^{-1/2} \right\}.
\eeq
The last term on the right-hand side is the tidal gravitational stress
corresponding to the $z$-component of the gravitational field of the central
point mass \cite{CK98}. Near the central sink, tidal squeezing of the disk
results in $Z/r = \sigma_{\rm n} / 2\rho_{\rm n} \ll 1$. So we may expand the last term and
find that equation (\ref{csazfb}) becomes
\beq
\rho_{\rm n}(r,t)C^2 = P_{\rm ext} + \frac{\pi}{2}G \sigma_{\rm n}^2(r,t) +
\frac{GM_{\rm cent}\sigma_{\rm n}^2}{8 \rho_{\rm n} r^3}.
\eeq
 
In order to follow shocks that might arise during the evolution,
tensor artificial viscosity \cite{TW79} is introduced. The
viscosity tensor $\overline{\overline{Q}}$ is given by
\beqar
\overline{\overline{Q}} & = & l^2 \rho_{\rm n} (\nabla \cdot \bvec{v})[\nabla\bvec{v}
- \frac{1}{3}(\nabla \cdot \bvec{v})\overline{\overline{I}}], \ \ \ \ \ \ \ if
\ \nabla \cdot \bvec{v} < 0, \nonumber \\
 & = & 0, \ \ \ \ \ \ \ \ \ \ \ \ \ \ \ \ \ \ \ \ \ \ \ \ \ \ \ \ \ \ \ \ \ \
 \ \ \ \ \ \ \ \ if \ \nabla
 \cdot  \bvec{v} \ge 0,  
\eeqar
where $l$ is a length comparable to the local mesh spacing, and
$\overline{\overline{I}}$ is the unit tensor. In regions
where the divergence of the velocity is negative, the radial viscosity force
(per unit volume) in axisymmetric geometry is given by (Fiedler \& Mouschovias 1992)
\beq
f_{Q,r} = - Q^i_{r;i} = -\frac{1}{r} \frac{\partial}{\partial r}(rQ^r_r)
-\frac{\partial Q^z_r}{\partial z} +  \frac{Q^{\phi}_{\phi}}{ r},      
\eeq
which, integrated over $z$ in the thin-disk approximation, becomes
\beqar
f_{Q,r} = & l^2 & \left[ -\frac{1}{3} \frac{v_{{\rm n},r}}{r} \frac{\partial
  \sigma_{\rm n}}{\partial r} \frac{\partial v_{{\rm n},r}}{\partial r}  -\frac{4}{3}
\frac{\sigma_{\rm n}}{r} \left( 
  \frac{\partial v_{{\rm n},r}}{\partial r}  \right)^2 \right. 
 -\frac{1}{3} \frac{\sigma_{\rm n} v_{{\rm n},r}}{r} \frac{\partial^2
   v_{{\rm n},r}}{\partial r^2} -\frac{2}{3} \frac{\partial\sigma_{\rm n}}{\partial r}
 \left( \frac{\partial v_{{\rm n},r}}{\partial r}  \right)^2 \nonumber \\ 
 & &  -\frac{4}{3} \sigma_{\rm n}  \frac{\partial v_{{\rm n},r}}{\partial r}
 \frac{\partial^2 v_{{\rm n},r}}{\partial r^2} + \frac{1}{3} \frac{\sigma_{\rm
     n}}{r^3}
 v_{{\rm n},r}^2 
\left. + \frac{1}{3} \frac{v_{{\rm n},r}^2}{r^2} \frac{\partial
     \sigma_{\rm n}}{\partial r}
 + \frac{\sigma_{\rm n}}{r^2} v_{{\rm n},r} \frac{\partial v_{{\rm n},r}}{\partial r}
  \right] .  \label{artvisc}
\eeqar
This term is added to the RHS of the radial force equation (\ref{thin5}).

\subsection{Initial and Boundary Conditions}

The calculation with the use of a central sink is considered a
continuation of the previous simulation of the formation and contraction of a
protostellar fragment in the following sense. When the central density exceeds
$\sim 10^{11}$ ${\rm cm^{-3}}$ and the
assumption of isothermality is no longer valid, the simulation is stoped and 
the central region of radius $R_{\rm inner}$ is removed from the active
computational mesh; it is now considered a sink cell. The calculation is then 
restarted, but the evolution is followed in detail only in the region $r \ge
R_{\rm inner}$,
in which the density is smaller than $10^{11}$ ${\rm cm^{-3}}$. Thus the
physical quantities at the end of the first calculation are used as initial
data for the new calculation.

At the inner boundary $R_{\rm inner}$ we impose perfectly
absorbing boundary conditions by
setting the values of all quantities at the boundary equal to those at the
first mass zone (beyond the boundary)

\begin{mathletters}
\hfill\begin{tabular}{ccc}
$v_{\rm n}(r=0) = v_{\rm n}(1)$,&
$v_{\rm e}(r=0) =v_{\rm e}(1)$,&
$v_{\rm i}(r=0) = v_{\rm i}(1)$,
\end{tabular}\hfill(\stepcounter{equation}\theequation,b,c)
\stepcounter{equation}\stepcounter{equation}

\hfill\begin{tabular}{ccc}
$v_{\rm g-}(r=0) = v_{\rm g-}(1)$,&
$v_{\rm g+}(r=0) = v_{\rm g+}(1)$,&
$v_{\rm g0}(r=0) = v_{\rm g0}(1)$,
\end{tabular}\hfill(\stepcounter{equation}\theequation,e,f)
\stepcounter{equation}\stepcounter{equation}
\end{mathletters}

\begin{mathletters}
\hfill\begin{tabular}{cc}
$\sigma_{\rm n}(r=0) = \sigma_{\rm n}(1)$,&
$\sigma_{\rm g-}(r=0) = \sigma_{\rm g-}(1)$,
\end{tabular}\hfill\stepcounter{equation}(\theequation,b)
\stepcounter{equation}

\hfill\begin{tabular}{cc}
$\sigma_{\rm g+}(r=0) = \sigma_{\rm g+}(1)$,&
$\sigma_{\rm g0}(r=0) = \sigma_{\rm g0}(1)$.
\end{tabular}\hfill\stepcounter{equation}(\theequation,d)
\stepcounter{equation}
\end{mathletters}

\noindent
(Ciolek \& K\"{o}nigl 1998 used linear extrapolation of these quantities from the first 
computational cell to the inner boundary.)
The spatial derivatives at the first computational zone are calculated
using one-sided differences instead of the central differencing used in the
rest of the computational grid. The outer-boundary conditions (at
$r=R$) are as in equations (\ref{out_bc1}) and (\ref{out_bc2}). 

\section{THE DIMENSIONLESS PROBLEM AND METHOD OF SOLUTION}

We put the equations in dimensionless form by choosing units 
natural to the model molecular clouds.  The unit of velocity is the isothermal sound speed in the
neutral fluid, $C$; the units of column density $\sigma_{\rm c,ref}$
and acceleration $2 \pi G \sigma_{\rm c,ref}$ are those of the neutral
column density and gravitational acceleration on the axis of
symmetry of
the reference state from which the initial equilibrium state is calculated.
The unit of magnetic field is its uniform value in the reference state, $B_{\rm ref}$. 
The units of all other quantities are derivable from these four. 

We denote the dimensionless form of the cylindrical polar coordinates
($r$, $z$) by ($\xi$, $\zeta$) and the dimensionless time by $\tau$,
so that equations (\ref{thin1})-(\ref{thin8}) become

\begin{mathletters}
\beq \label{dimless1}
\sigma_{\rm n}(\xi,\tau) =  2\rho_{\rm n}(\xi,\tau)Z(\xi,\tau), 
\eeq
\beq
\frac{\partial \sigma_{\rm n}}{\partial \tau} + \frac{1}{\xi}\frac{\partial (\xi 
\sigma_{\rm n} v_{{\rm n},\xi})}{\partial \xi}  =  0, 
\eeq
\beq
\frac{\partial (\chi_{\rm g} \sigma_{\rm n})}{\partial \tau} + \frac{1}{\xi}\frac{\partial [\xi 
\sigma_{\rm n} (\chi_{\rm g-} v_{{\rm g-},\xi} + \chi_{\rm g+} v_{{\rm
    g+},\xi} + \chi_{\rm g0}
v_{{\rm g0},\xi})]}{\partial \xi}  =  0, 
\eeq
\beq
\rho_{\rm n}(\xi,\tau)=\frac{1}{4} \left[\tilde{P}_{\rm ext} + \sigma_{\rm n}^2(\xi,\tau)\right],
\eeq
\beq
\frac{\partial (\sigma_{\rm n} v_{{\rm n},\xi})}{\partial \tau} + \frac{1}{\xi}\frac{\partial( \xi 
\sigma_{\rm n} v_{{\rm n},\xi}^2)}{\partial \xi}  =  -\tilde{C}_{\rm eff}^2 \frac{\partial
\sigma_{\rm n}}{\partial \xi} + \sigma_{\rm n} g_{\xi} + F_{{\rm mag},\xi} ,
\eeq
\beq
\tilde{C}_{\rm eff}^2 = \sigma_{\rm n}^2 \frac{(3\tilde{P}_{\rm ext} +
  \sigma_{\rm n}^2)}{(\tilde{P}_{\rm ext} + \sigma_{\rm n}^2)^2},
\eeq
\beq
F_{{\rm mag},\xi} = \frac{1}{\mu_{\rm d,c0}^2} \left\{ B_{\zeta,{\rm eq}}\left( B_{\xi,Z}-Z\frac{\partial
  B_{\zeta,{\rm eq}}}{\partial \xi}\right) + \frac{1}{2}
\frac{\partial Z}{\partial \xi} \left[B_{\xi,Z}^2
+ 2B_{\zeta,{\rm eq}}\left(B_{\xi,Z}\frac{\partial Z}{\partial \xi}\right)+\left(B_{\xi,Z}\frac{\partial
  Z}{\partial \xi}\right)^2\right] \right\}, 
\eeq
\beq
g_{\rm \xi}(\xi) = \int_{0}^{\infty}d\xi' \xi' \sigma_{\rm n}(\xi')\mathcal{M}(\xi,\xi'),
\eeq
\beq
\mathcal{M}(\xi,\xi') = \frac{2}{\pi}\frac{d}{d\xi}\left[ \frac{1}{\xi_>}K\left(
  \frac{\xi_<}{\xi_>}\right)  \right],
\eeq
\beq
B_{\xi,Z}(\xi) = - \int_{0}^{\infty}d\xi' \xi' [B_{\zeta,{\rm eq}}(\xi')-B_{\rm ref}]\mathcal{M}(\xi,\xi'),
\eeq
\beq
\frac{\partial B_{\zeta,{\rm eq}}}{\partial \tau} + \frac{1}{\xi}\frac{\partial (\xi 
B_{\zeta,{\rm eq}} v_{{\rm e},\xi})}{\partial \xi}  =  0, 
\eeq
\beq
\ve = \vn + \frac{F_{\rm mag}\tne}{\sigma_{\rm n}\Lambda_{\rm e}} \,.
\eeq
\end{mathletters}
%%%%%%%%%%%%%%%%%%%%%%%%%%%%%%%%%%%%%%%%%%%%%%%%%%%%%%%%%%%%%%%%%%%%%%%%%%%%%%%%%%%%%%%

\noindent
Equations (\ref{thevels}l-t) are already in
dimensionless form, provided that the velocities used are in units of $C$.
After the introduction of the central sink, equations
(\ref{csa1})-(\ref{grav_field}), (\ref{csazfb}) and (\ref{artvisc}) become
in dimenssionless form:

\begin{mathletters}
\beq \label{dimless2}
\left. \frac{\partial M_{*}}{\partial \tau} = - \xi \sigma_{\rm n} v_{\rm
  n} \right|_{\xi = \Xi_{\rm inner}}, 
\eeq
\beq
\left. \frac{\partial M_{\rm g,*}}{\partial \tau} = -\xi (\sigma_{\rm g-} v_{\rm g-}+
\sigma_{\rm g+}v_{\rm g+} + \sigma_{\rm g0}v_{\rm g0}) \right|_{\xi=\Xi_{\rm inner}}, 
\eeq
\beq
\left. \frac{\partial \Phi_{\rm B,*}}{\partial \tau} = -2 \xi 
B_{\zeta,{\rm eq}}v_{\rm e} \right|_{\xi=\Xi_{\rm inner}}, 
\eeq
\beq
g_\xi(\xi) = -\frac{M_{*}}{\xi^2} + \int_{0}^{\infty}d\xi' \xi'
\sigma_n(\xi')\mathcal{M}(\xi,\xi'),
\eeq
\beq
\rho_{\rm n}(\xi,\tau)=\frac{1}{8} (\tilde{P}_{\rm ext} + \sigma_{\rm
  n}^2)\left\{1+\left[1+\frac{8M_{*}\sigma_{\rm n}^2}{\xi^3(\tilde{P}_{\rm ext} + \sigma_{\rm 
    n}^2)^2}\right]^{1/2}\right\} ,
\eeq
\beqar
f_{Q,\xi} = & \tilde{l}^2 & \left[ -\frac{1}{3} \frac{v_{\rm n}}{\xi} \frac{\partial
  \sigma_{\rm n}}{\partial \xi} \frac{\partial v_{\rm n}}{\partial \xi}  -\frac{4}{3}
\frac{\sigma_{\rm n}}{\xi} \left( 
  \frac{\partial v_{\rm n}}{\partial \xi}  \right)^2 \right. 
 -\frac{1}{3} \frac{\sigma_{\rm n} v_{\rm n}}{\xi} \frac{\partial^2
   v_{\rm n}}{\partial \xi^2} -\frac{2}{3} \frac{\partial\sigma_{\rm n}}{\partial \xi}
 \left( \frac{\partial v_{\rm n}}{\partial \xi}  \right)^2 \nonumber \\ 
 & &  -\frac{4}{3} \sigma_{\rm n}  \frac{\partial v_{\rm n}}{\partial \xi}
 \frac{\partial^2 v_{\rm n}}{\partial \xi^2} + \frac{1}{3} \frac{\sigma_{\rm
     n}}{\xi^3}
 v_{\rm n}^2 
\left. + \frac{1}{3} \frac{v_{\rm n}^2}{\xi^2} \frac{\partial
     \sigma_{\rm n}}{\partial \xi}
 + \frac{\sigma_{\rm n}}{\xi^2} v_{\rm n} \frac{\partial v_{\rm n}}{\partial \xi}
  \right] .  
\eeqar
\end{mathletters}

\noindent
Equations (\ref{dimless1}-l), (\ref{thevels}l-t) and 
(\ref{dimless2}-f) comprise the dimensionless system to be
solved numerically. The dependent variables are now
dimensionless but in this section only, for ease in identification, we
have retained the same symbols as for their dimensional counterparts.
The quantities $\Theta_s$ and $\Xi_{\phi s}$ in the expressions for
the velocities are already dimensionless (see Appendix B). The free
parameters of the problem are: 
the initial central mass-to-flux ratio
$\mu_{\rm d,c0}=(2\pi G^{1/2}\sigma_{\rm c,ref})/B_{ref}$ in units of its
critical value for gravitational collapse in disk geometry,
$1/(2\pi G^{1/2})$ (Nakano \& Nakamura 1978);
and the constant external
pressure relative to the central gravitational ``pressure'' along the
axis of symmetry of
the reference state $\tilde{P}_{\rm ext} \equiv P_{\rm ext}/
(\frac{\pi}{2}G\sigma_{\rm c,ref})$.  
The reaction rates of the chemical network and the value of the radius of the
dust grains are given in Appendix A. [The thermal (Jeans) length does
  not appear explicitly in the equations because of
the choice of the units we are using: The unit of length
$C^2 / 2 \pi G \sigma_{\rm c,ref}$ is proportional to the Jeans length.]

A control-volume method is employed for solving the dimensionless
equations that govern the evolution of the model cloud. The method consists of
an implicit time integrator; a conservative upwind advective difference scheme
that incorporates the second-order accurate monotonicity algorithm of Van
Leer (1979); a second-order difference approximation of forces inside a mass
zone; an integral approximation of the gravitational and magnetic fields; and
an adaptive mesh capable of resolving the smallest natural lengthscale
(i.e., thermal lengthscale $\lambda_T$). The algebraic equations
for the equilibrium abundances of charged species are solved at each time
step. A detailed description of the numerical methods is given by Morton,
Mouschovias \& Ciolek (1994). Once a central sink is introduced, we keep the grid 
stationary but nonuniform.

\section{DISCUSSION}
 
We have formulated the problem of the late accretion phase of a
magnetic, isothermal disk surrounding a forming protostar. Our formulation 
is such that we can follow the structure and evolution of the disk after 
the formation of an opaque core at the center of the disk, but 
 before the protostar turns on and radiation from it affects the surrounding 
matter. The
central-sink approximation allows us to follow the 
evolution of the system at significantly later times without the need of 
a numerically costly radiative transfer calculation for the opaque core. 

This is a similar approach to that of Ciolek and
K\"{onigl} (1998) with two major physical and one numerical differences: (i) We
have accounted for an additional charged species (positively-charged
grains, which become important above densities $\sim 10^{9}$ ${\rm
  cm^{-3}}$). (ii) We have assumed that the magnetic flux is 
frozen only in the electron fluid and we have allowed the physics of the
problem to determine the degree of attachment of the ions to the magnetic
field lines. Detachment of
ions in fact begins at densities $\sim 10^8$ ${\rm cm ^{-3}}$, as
shown by Desch \& Mouschovias (2001). (iii) We 
have used a more accurate calculation of the
magnetic field and more appropriate boundary conditions. We have also included 
artificial viscosity in our code in order to resolve and track the 
formation and propagation of shocks. These differences result in 
new physical phenomena, such as the quasi-periodic formation and dissipation of
shocks.
The results of our numerical calculation are described in Paper II.

\acknowledgements{We thank Glenn Ciolek, Chester Eng and Vasiliki Pavlidou
for useful discussions. This work was carried out without external support, 
and would not have been published without the generosity of the ApJ, and the
Astronomy and Physics Departments of the University of Illinois. KT was 
supported by the University of Illinois Research Board and the Greek State 
Scholarship Foundation during this project.}

%%%%%%%%%%%%%%%%%%%%%%%%%%%%%%%%%%%%%%%%%%%%%%%%%%%%%%%
% end of main text
%%%%%%%%%%%%%%%%%%%%%%%%%%%%%%%%%%%%%%%%%%%%%%%%%%%%%%%
\appendix
\section{ABUNDANCES OF RELEVANT SPECIES}

The abundances of all relevant species (electrons, molecular ions,
atomic ions, charged and neutral grains) are determined by assuming that chemical
equilibrium is established and
maintained throughout the evolution of the model molecular clouds. 
This assumption is valid since the chemical-reaction timescales are
smaller than the evolutionary timescales of the model cores for
neutral densities up to $10^{14}$ ${\rm cm^{-3}}$.

In equilibrium, the rate of production of a specific species is equal
to the rate of its destruction through the relevant chemical reactions listed
in Table \ref{table1}. Then the relative
abundances of all species can be found solving the system of
the creation/destruction rate equations for electrons, molecular ions,
atomic ions, and positive grains:
\beqar
\zeta_{\rm CR}n_{\rm n}& = &(\alpha_{\rm dr}\chi_{\rm m_+} + \alpha_{\rm
  rr}\chi_{\rm A^+} +
\alpha_{\rm eg^0}\chi_{\rm g^0} + \alpha_{\rm eg^+}\chi_{\rm g^+})\chi_{\rm e}
n_{\rm n}^2, \label{eqa1}\\
\zeta_{\rm CR}n_{\rm n}& = &(\alpha_{\rm dr}\chi_{\rm e} + \beta\chi_{\rm A^0} +
\alpha_{\rm m_+g^-}\chi_{\rm g^-} + \alpha_{\rm m_+g^0}\chi_{\rm
  g^0})\chi_{\rm m_+} n_{\rm n}^2,\label{eqa2}\\
\beta \chi_{\rm A^0} \chi_{\rm m_+} n_{\rm n}^2 & = & [\alpha_{\rm rr}\chi_{\rm e}
+ \alpha_{\rm A^+g^-}\chi_{\rm g^-} + \alpha_{\rm A^+g^0}\chi_{\rm
  g^0}]\chi_{\rm A^+} n_{\rm n}^2,\label{eqa3}\\
(\alpha_{\rm A^+g^0}\chi_{\rm A^+} + \alpha_{\rm m_+g^0}\chi_{\rm m_+})
  \chi_{\rm g^0} n_{\rm n}^2 & =& (\alpha_{\rm eg^+}\chi_{\rm e} + \alpha_{\rm g^+g^-}\chi_{\rm g^-})\chi_{\rm
  g^+} n_{\rm n}^2 ,\label{eqa4}\\
\chi_{\rm g} & = & \chi_{\rm g^-} + \chi_{\rm g^+} + \chi_{\rm g^0}. \label{eqa5}
\eeqar
Since the total relative abundance of grains $\chi_{\rm g}$ and the neutral
density $n_{\rm n}$ are obtained at each timestep from the continuity
equations (31b) and (31c), the abundances of
both negative and neutral grains can be determined from the system
(\ref{eqa1})-(\ref{eqa5}) once local charge neutrality is imposed:
\beq
0 = \chi_{\rm m_+} + \chi_{\rm A^+} +  \chi_{\rm g^+} - \chi_{\rm e} - \chi_{\rm
  g^-}. \label{eqa6} 
\eeq
The relative abundance of species $s$ is denoted by $\chi_s(\equiv
n_s/n_{\rm n})$.

The values of the rate constants for the gas phase
reactions are taken from Ciolek \& Mouschovias (1995) and
references therein. For the dissociative recombination of
${\rm HCO^+}$, $\alpha_{\rm dr} = 10^{-6}$ ${\rm cm^3}$ ${\rm s^{-1}}$; for the radiative recombination of
atomic 
ions, $\alpha_{\rm rr} = 10^{-11}$ ${\rm cm^3}$ ${\rm s^{-1}}$; and for the charge
exchange between atomic and  
molecular ions, $\beta = 1.1 \times 10^{-9}$ ${\rm cm^3}$ ${\rm
  s^{-1}}$. The typical value of the
cosmic-ray ionization rate 
is $\zeta_{CR} = 5 \times 10^{-17}$ ${\rm s^{-1}}$.

The rate constants for electron-grain and ion-grain
collisions (in ${\rm cm^3}$ ${\rm s^{-1}}$) were originally 
calculated by Spitzer (1941, 1947) and later modified by Draine and Sutin (1987)
to include corrections due to polarization of a grain by the approaching
charged particle:
\beq \label{eqa7}
\alpha_{\rm eg^0} =  \pi a^2 \sqrt{\frac{8k_BT}{\pi m_{\rm e}}} \left\{ 1 +
    \sqrt{\frac{\pi e^2}{2 a k_{\rm B} T}}  \right\} \mathcal{P}_{\rm e} ,
\eeq

\beq
\alpha_{\rm ig^0} =  \pi a^2 \sqrt{\frac{8k_{\rm B}T}{\pi m_{\rm i}}} \left\{ 1 +
    \sqrt{\frac{\pi e^2}{2 a k_{\rm B} T}}  \right\} \mathcal{P}_{\rm i} ,
\eeq

\beq
\alpha_{\rm eg^+} =  \pi a^2 \sqrt{\frac{8k_{\rm B}T}{\pi m_{\rm e}}} \left[1 +
\frac{e^2}{a k_{\rm B} T} \right] \left\{ 1 +
    \sqrt{\frac{2}{2 + (a k_{\rm B} T/e^2)}}  \right\} \mathcal{P}_{\rm e} ,
\eeq

\beq
\alpha_{\rm ig^-} =  \pi a^2 \sqrt{\frac{8k_{\rm B}T}{\pi m_{\rm i}}} \left[1 +
\frac{e^2}{a k_{\rm B} T} \right] \left\{ 1 +
    \sqrt{\frac{2}{2 + (a k_{\rm B} T/e^2)}}  \right\} \mathcal{P}_{\rm i}.
\eeq
The quantities $\alpha_{\rm eg^0}$, $\alpha_{\rm eg^+}$ are the rates of capture of $e^-$
onto neutral and positive grains, respectively; and $\alpha_{\rm
  ig^0}$ and 
$\alpha_{\rm ig^-}$ are the rates of attachment of ions onto neutral and negative
grains, respectively. Similarly, the rate of charge transfer and neutralization
by direct grain-grain collisions is
\beq \label{eqa11}
\alpha_{\rm g^-g^+} =  \pi (a_1+a_2)^2 \sqrt{\frac{8k_{\rm B}T}{\pi m_{\rm
      red}}} \left[1 + \frac{e^2}{\alpha k_{\rm B} T} \right] \left\{ 1 +
    \sqrt{\frac{2}{2 + (\alpha k_{\rm B} T/e^2)}}  \right\}.
\eeq
In equations (\ref{eqa7})-(\ref{eqa11}) $e$ is the charge of the electron, $m_{\rm e}$ and $m_{\rm i}$ the
masses of the electron and ion, respectively, $k_{\rm B}$ the Boltzmann constant,
$T$ the temperature, $\mathcal{P}_{\rm e}(\mathcal{P}_{\rm i})$ the sticking probability for an
electron (ion) onto a grain, and $m_{\rm red} = m_1m_2/(m_1+m_2)$ the reduced mass of the two grains.
The terms enclosed in curly brackets account for
the polarization of a grain by the incoming charge.

The grains are
assumed spherical with radius $a = 3.75 \times 10^{-6}$ cm and
density $\rho_{\rm g} = 2.3$ g ${\rm cm^{-3}}$, the average density of
silicates. The 
temperature is kept at the constant value $T = 10$ K. The mass of
molecular ions is that of HCO, $m_{\rm m+} = 29$
$m_{\rm p}$, while for atomic ions we use a mean value $m_{\rm A} = 23.5$
$m_{\rm p}$, in-between
the mass of Na ($m_{\rm Na} = 23$ $m_{\rm p}$) and the mass of Mg ($m_{\rm Mg}
= 24$ 
$m_{\rm p}$). Since the masses of all the dominant ions are comparable, the fact that different
ionic species dominate in different density regimes does not affect the
evolution of the cloud cores. Finally, the sticking probability
of electrons and ions onto grains are taken to be $\mathcal{P}_{\rm e} =
0.6$ and $\mathcal{P}_{\rm i} = 1$, respectively \citep{DraineSutin87, UN80, DM01}.

%%%%%%%%%%%%%%%%%%%%%%%%%%%%%%%%%%%%%%%%%%%%%%%%%%%%%%%%%
% second appendix here
%%%%%%%%%%%%%%%%%%%%%%%%%%%%%%%%%%%%%%%%%%%%%%%%%%%%%%%%
\section{SPECIES VELOCITIES IN TERMS OF THE ELECTRON AND NEUTRAL VELOCITIES} 

Once the velocity of the neutrals $v_n$ is known, the
velocities of the other species can be found by solving the linear
algebraic system of the force equations of the species
and Amp\`{e}re's law (eqs. [\ref{eforceq}]-[\ref{g0forceq}] and [\ref{amp}],
where $\bvec{j}$ in eq. [\ref{amp}] is given by eq. [\ref{jdef}]).

We add together equations (\ref{eforceq})-(\ref{g0forceq}) and we use equation (\ref{amp}) to
eliminate $\bvec{j}$ in the resulting equation to find the {\em
  plasma force equation}:
\beq \label{plasma}
\frac{\rn}{\tne}(\bvec{v_{\rm e}}-\bvec{v_{\rm n}})+
\frac{\rn}{\tni}(\bvec{v_{\rm i}}-\bvec{v_{\rm n}})+
\frac{\rn}{\tngm}(\bvec{v_{\rm g-}}-\bvec{v_{\rm n}})+
\frac{\rn}{\tngp}(\bvec{v_{\rm g+}}-\bvec{v_{\rm n}})+
\frac{\rn}{\tngo}(\bvec{v_{\rm g0}}-\bvec{v_{\rm n}})=
\frac{(\nabla\times\bvec{B})\times \bvec{B}}{4\pi}\,.
\eeq
The terms on the left-hand side are the collisional coupling (momentum
exchange) forces between the neutrals and the other five species
(e, i, ${\rm g_-}$, ${\rm g_+}$ and  ${\rm g_0}$). They were present
on the right-hand side of the force equation (\ref{nforceq}) of the neutrals, but
were eliminated by using equation (\ref{plasma}), which is a
quantitative description of why and how the magnetic force appears in
the force equation of the neutrals. 

We now use equation (\ref{fluxfreezing}) for the electric field to
eliminate $\bvec{E}$ from equations
(\ref{iforceq})-(\ref{gpforceq}) and find that
\beqar
0&=&\frac{en_{\rm i}}{c}(\bvec{v_{\rm i}}-\bvec{v_{\rm e}})\times \bvec{B} 
+ \frac{\ri}{\tin}(\bvec{v_{\rm n}}-\bvec{v_{\rm i}})\label{isimple}\\
0&=&-\frac{e\ngm}{c}(\bvec{v_{\rm g-}}-\bvec{v_{\rm e}})\times \bvec{B} 
+ \frac{\rgm}{\tgn}(\bvec{v_{\rm n}}-\bvec{v_{\rm g-}})
+\frac{\rgo}{\tau_-}(\bvec{v_{\rm g0}}-\bvec{v_{\rm g-}})
\label{msimple}
\\
0&=&\frac{en_{\rm g+}}{c}(\bvec{v_{\rm g+}}-\bvec{v_{\rm e}})\times \bvec{B} 
+ \frac{\rho_{\rm g+}}{\tgn}(\bvec{v_{\rm n}}-\bvec{v_{\rm g+}})
+\frac{\rgo}{\tau_+}(\bvec{v_{\rm g0}}-\bvec{v_{\rm g+}})
\label{psimple}
\eeqar
where 
\beqar
\tau_- &=& \left(
\frac{1}{\tau_{\rm g0e, inel}}+
\frac{\rgm}{\rgo}\frac{1}{\tau_{\rm g-i, inel}} + 
\frac{\rgm}{\rgo}\frac{1}{\tau_{\rm g-g+, inel}}\right)^{-1}
\\
\tau_+ &=& \left(
\frac{1}{\tau_{\rm g0i, inel}}+
\frac{\rgp}{\rgo}\frac{1}{\tau_{\rm g+e, inel}} + 
\frac{\rgp}{\rgo}\frac{1}{\tau_{\rm g+g-, inel}}\right)^{-1}
\eeqar
are the relevant timescales for inelastic collisions between grains.
In particular, $\tau_-$ is the timescale for a neutral grain to be
involved in {\em any} inelastic reaction that converts
negative grains to neutral grains or vice versa, and $\tau_+$ is the
timescale for a neutral grain to participate in any inelastic reaction
that transforms positive and neutral grains into each other. 

Equations (\ref{plasma})-(\ref{psimple}), together with the neutral-grain
force equation, 
\beq\label{nsimple}
0= \frac{\rgo}{\tgn}(\bvec{v_{\rm n}}-\bvec{v_{\rm g0}})
+\frac{\rgo}{\tau_+}(\bvec{v_{\rm g+}}-\bvec{v_{\rm g0}})
+\frac{\rgo}{\tau_-}(\bvec{v_{\rm g-}}-\bvec{v_{\rm g0}})\,\,,
\eeq
form a $10\times 10$ linear system of equations, with unknowns being the
$r-$ and $\phi-$ components of the velocities of electrons, ions, and 
positive, negative, and neutral grains. 

We now integrate equations (\ref{plasma})-(\ref{psimple}) and (\ref{nsimple})
in the $z$-direction using the
one-zone approximation (no variation of quantities with $z$ inside the
disk) and decompose them to $r$- and $\phi$-components
(assuming that there is no bulk rotation of the disk, $v_{n,\phi}=0$). Note
that {\it ratios} of mass densities, $\rho$, 
equal ratios of column densities, $\sigma$, and the former rather than
the latter are used whenever such a ratio appears. The resulting
$10 \times 10$ linear system (with unknowns $\vi, \vif, \ve, \vef, \vgm,
\vgmf, \vgp, \vgpf, \vgo, \vgof$) is:
\beqar
0&=& \oitin(\vif-\vef) + \vn -\vi \label{eqir}\\
0&=& \oitin(\ve-\vi) -\vif \label{eqif}\\
0&=& \ogtgn (\vef-\vgmf)+\vn -\vgm +\rmm(\vgo-\vgm) \label{eqmr}\\
0&=& \ogtgn(\vgm - \ve) -\vgmf + \rmm(\vgof-\vgmf)\label{eqmf}\\
0&=& \ogtgn(\vgpf - \vef) + \vn - \vgp + \rpp(\vgo-\vgp)\label{eqpr}\\
0&=& \ogtgn(\ve-\vgp) -\vgpf + \rpp(\vgof-\vgpf)\label{eqpf}\\
0&=& \vn - \vgo + \frac{\tgn}{\tm}(\vgm-\vgo) + \frac{\tgn}{\tp}(\vgp-\vgo)\label{eq0r}\\
0&=& - \vgof + \frac{\tgn}{\tm}(\vgmf-\vgof) + \frac{\tgn}{\tp}(\vgpf-\vgof)\label{eq0f}\\
0&=& \vef+\frac{\tne}{\tni}\vif + \frac{\tne}{\tngp}\vgpf 
+\frac{\tne}{\tngm}\vgmf + \frac{\tne}{\tngo}\vgof\label{eqplf}\\
\frac{F_{\rm mag}\tne}{\sigma_{\rm n}}&=& 
\ve - \vn + \frac{\tne}{\tni}(\vi-\vn) + \frac{\tne}{\tngm}(\vgm-\vn)
+ \frac{\tne}{\tngp}(\vgp - \vn) + \frac{\tne}{\tngo}(\vgo-\vn)\nonumber\label{eqplr}\\
\eeqar

We first solve the system of equations (\ref{eqir}) - (\ref{eqplf}) in
that we express all the velocities in terms of
$\vn$ and $\ve$, and then substitute the resulting expressions for all velocities in equation (\ref{eqplr})
to find $\ve$ as a function of $\vn$. We note that in matrix form, the
system of (\ref{eqir}) - (\ref{eqplf}) is written as
\beq
{\cal A}\bvec{V}= \vn \bvec{C^{({\rm n})}} + \ve \bvec{C^{({\rm e})}}\,,
\eeq
where $\cal A$ is the $9 \times 9$ matrix of coefficients
\beq
{\cal A} = \left[\begin{array}{ccccccccc}
-1 & 0 & 0& 0& \oitin& 0& 0& 0& -\oitin \\ 
-\oitin & 0 & 0 & 0 & -1 & 0& 0& 0& 0\\
0& -1-r_- & 0& r_- &0 &-\ogtgn & 0&0& \ogtgn\\
0&\ogtgn & 0&0&0& -1-r_- & 0&r_-&0\\
0&0& -1-r_+ & r_+&0&0&\ogtgn& 0 & -\ogtgn \\
0&0&-\ogtgn&0& 0&0&-1-r_+ & r_+ & 0 \\
0&\displaystyle \frac{\tgn}{\tm}&\displaystyle \frac{\tgn}{\tp} & 
-\displaystyle \frac{\tgn}{\tz} & 0 & 0&0&0&0\\
0&0&0& 0&0&\displaystyle \frac{\tgn}{\tm} &\displaystyle \frac{\tgn}{\tp} & 
\displaystyle -\frac{\tgn}{\tz}& 0 \\
0&0&0&0&\displaystyle \frac{\tne}{\tni} &\displaystyle \frac{\tne}{\tngm} &\displaystyle \frac{\tne}{\tngp}&
\displaystyle \frac{\tne}{\tngo} & 1 
\end{array}\right],
\eeq
with $\tau_0 =
(1 / \tau_{gn}+ 1/\tau_- + 1 / \tau_+)^{-1}$,
$r_- = \rho_{\rm g0} \tau_{\rm gn}/(\rho_{\rm g-} \tau_{-})$, $r_+ =
\rho_{\rm g0} \tau_{\rm gn}/(\rho_{\rm g+} \tau_{+}) $, 
and the three column vectors $\bvec{V}$, $\bvec{C^{(n)}}$ and $\bvec{C^{(e)}}$ are defined by
\beq
\begin{array}{ccc}
\bvec{V} = \left[ \begin{array}{c}
\vi \\\vgm\\\vgp\\\vgo\\\vif \\ \vgmf\\ \vgpf \\ \vgof \\ \vef
\end{array} \right] \,\,,&
\bvec{C^{({\rm n})}} = \left[ \begin{array}{r}
-1 \\0\\-1\\0\\-1 \\0\\-1\\0 \\0
\end{array} \right] \,\,,&
 \bvec{C^{({\rm e})}} = \left[ \begin{array}{c}
0 \\-\oitin\\0\\\ogtgn\\0 \\-\ogtgn\\0\\0 \\0
\end{array} \right]
\end{array}\,.
\eeq

We use Cramer's method to solve the above system. We define 
\beq
D=det[{\cal A}]\,.
\eeq
The notation $D^{\rm n}_s$ represents the
determinant of $\cal A$ in which the column corresponding to the
coefficients in front of the velocity of species $s$ (where $s$ is one
of the following: i; ${\rm i_{\phi}}$; ${\rm g_{-}}$; ${\rm g_{-\phi}}$; ${\rm g_{+}}$;
${\rm g_{+\phi}}$; ${\rm g_{0}}$; ${\rm g_{0\phi}}$; ${\rm e_{\phi}}$) has been 
substituted by $\bvec{C^{({\rm n})}}$. Similarly, $D^{\rm e}_s$ denotes the
determinant of $\cal A$ in which the column $s$ has been replaced by
$\bvec{C^{({\rm e})}}$. Then, the solution of the system is: 

\begin{mathletters}
\hfill\begin{tabular}{ll}
$\displaystyle \vi = \frac{D^{\rm n}_{\rm i}}{D}\vn + \frac{D^{\rm e}_{\rm i}}{D}\ve$,&
$\displaystyle \vif = \frac{D^{\rm n}_{{\rm i},\phi}}{D}\vn + \frac{D^{\rm e}_{{\rm i},\phi}}{D}\ve$,
\end{tabular}\hfill(\stepcounter{equation}\theequation,b)
\stepcounter{equation}

\hfill\begin{tabular}{ll}
$\displaystyle \vgm = \frac{D^{\rm n}_{\rm g-}}{D}\vn + \frac{D^{\rm e}_{\rm g-}}{D}\ve$,&
$\displaystyle \vgmf = \frac{D^{\rm n}_{{\rm g-},\phi}}{D}\vn + \frac{D^{\rm e}_{{\rm g-},\phi}}{D}\ve$,
\end{tabular}\hfill(\stepcounter{equation}\theequation,d)
\stepcounter{equation}

\hfill\begin{tabular}{ll}
$\displaystyle \vgp = \frac{D^{\rm n}_{\rm g+}}{D}\vn + \frac{D^{\rm e}_{\rm g+}}{D}\ve $,&
$\displaystyle \vgpf = \frac{D^{\rm n}_{{\rm g+},\phi}}{D}\vn + \frac{D^{\rm e}_{{\rm g+},\phi}}{D}\ve $,
\end{tabular}\hfill(\stepcounter{equation}\theequation,f)
\stepcounter{equation}

\hfill\begin{tabular}{ll}
$\displaystyle \vgo = \frac{D^{\rm n}_{\rm g0}}{D}\vn + \frac{D^{\rm e}_{\rm g0}}{D}\ve $,&
$\displaystyle \vgof = \frac{D^{\rm n}_{{\rm g0},\phi}}{D}\vn + \frac{D^{\rm e}_{{\rm g0},\phi}}{D}\ve $,
\end{tabular}\hfill(\stepcounter{equation}\theequation,h)
\stepcounter{equation}

\hfill\begin{tabular}{l}
$\displaystyle \vef = \frac{D^{\rm n}_{{\rm e},\phi}}{D}\vn + \frac{D^{\rm e}_{{\rm e},\phi}}{D}\ve$.
\end{tabular}\hfill(\stepcounter{equation}\theequation)
\end{mathletters}

\noindent
Note that for all $r$-components of the velocities, $D^{\rm e}_s+D^{\rm n}_s =D$. 
This can easily be seen if in the determinant $D^{\rm e}_s+D^{\rm n}_s$ (which 
is the same as the determinant obtained by substituting the
$s$-column by $\bvec{C^{\rm e}}+\bvec{C^{\rm n}}$) we add to the s-column the 3
{\em other} columns that also correspond to $r$-velocities. The result
is always the original determinant of the coefficients, $D$. This is a
consequence of the property of the system of equations (\ref{eqir}) --
(\ref{eqplr})
that $r$-velocities, whether known or unknown, always appear as velocity
{\em differences} in the equations. This is {\em not} the case for the 
$\phi$-components of the velocities, where the assumption
$v_{{\rm n},\phi}=0$ breaks the corresponding symmetry. 

This result is the mathematical expression of a physical property of
the $r$-velocities, namely, that they vary smoothly between the $r$-velocity of the
electrons, which are attached to the field lines, and the 
$r$-velocity of the neutrals. When a species is well attached to the
field lines, its $r$-velocity coincides with that of the
electrons. When collisions with the neutrals become frequent enough
that the species is fully detached from the field lines, its $r$-velocity
becomes equal to that of the neutrals. This physical picture can
further be clarified by defining the quantities
$\Xi_s = (D^{\rm e}_s+D^{\rm n}_s)/D$ and $\Theta_s = D^{\rm e}/D^{\rm n}$, and noting
that, as explained above, $\Xi_s=1$ for all $r$-velocities. We can
then rewrite the solution of the system as in equations
(\ref{thevels}a) -- (\ref{thevels}i). 
For the $r$-velocities $\Theta_s$ is then the {\em attachment
  parameter} (i.e. for $\Theta_s \gg 1$, $v_s \approx \ve$, while for
$\Theta_s \ll 1$, $v_s \approx \vn$).

Finally, $\ve$ can be obtained as a function of $\vn$ by substituting
equations (\ref{thevels}a) -- (\ref{thevels}i) in equation (\ref{eqplr}), which becomes
\beq
\frac{F_{\rm mag}\tne}{\sigma_{\rm n}}=
(\ve - \vn)\left(1+
\frac{\tne}{\tni}\frac{\Theta_{\rm i}}{\Theta_{\rm i}+1}+
\frac{\tne}{\tngm}\frac{\Theta_{\rm g-}}{\Theta_{\rm g-}+1}+
\frac{\tne}{\tngp}\frac{\Theta_{\rm g+}}{\Theta_{\rm g+}+1}+
\frac{\tne}{\tngo}\frac{\Theta_{\rm g0}}{\Theta_{\rm g0}+1}\right)\,.
\eeq
Defining 
\beq
\Lambda_{\rm e} = 1+
\frac{\tne}{\tni}\frac{\Theta_{\rm i}}{\Theta_{\rm i}+1}+
\frac{\tne}{\tngm}\frac{\Theta_{\rm g-}}{\Theta_{\rm g-}+1}+
\frac{\tne}{\tngp}\frac{\Theta_{\rm g+}}{\Theta_{\rm g+}+1}+
\frac{\tne}{\tngo}\frac{\Theta_{\rm g0}}{\Theta_{\rm g0}+1}\,\,,
\eeq
we get equation (\ref{theevel})
which, if substituted in equations (\ref{thevels}a) -- (\ref{thevels}i) 
gives the
remaining $r$-velocities and the $\phi$-velocities of all species as 
functions of $\vn$.

%%%%%%%%%%%%%%%%%%%%%%%%%% TABLE 1 %%%%%%%%%%%%%%%%%%%%%%%%%%%%%%%%%%%%%

\begin{table}
\begin{center}
\caption{\label{table1} Chemical reaction network used in the
  calculation of the abundances of charged species.}
\begin{tabular}{l c c c}
\tableline\tableline
\multicolumn{4}{c}{Relevant Chemical Reactions in Molecular Clouds}\\
\tableline
Cosmic-Ray Ionization:            & ${\rm H_2 + CR}$ & $\rightarrow$ & ${\rm H_2^+ + e}$ \\
                                  & ${\rm H_2^+ +  H_2}$ & $\rightarrow$ & ${\rm H_3^+ +  H} $\\
                                  & ${\rm H_3^+ + CO}$ & $\rightarrow$ & ${\rm HCO^+ + H_2}$ \\
Dissociative Recombination:       & ${\rm HCO^{+} + e}$ & $\rightarrow$ &${\rm H + CO} $\\
Radiative recombination:          & ${\rm A^{+} + e}$ & $\rightarrow$ & ${\rm A^{0}} + h{\rm \nu}$ \\
Charge tranfer:                   & ${\rm A^0 + HCO^{+}}$ & $\rightarrow$ &${\rm A^{+} + HCO}$ \\
$e^-$ attachment onto grains:     & ${\rm e + g_{0}}$ & $\rightarrow$ & ${\rm g_{-}} $  \\
                                  & ${\rm e + g_{+}}$ & $\rightarrow$ & ${\rm g_{0}} $  \\
Atomic-ion attachment onto grains:& ${\rm A^{+} + g_{-}}$ & $\rightarrow$ & ${\rm A^0 + g_{0}} $   \\
                                  & ${\rm A^{+} + g_{0}}$ & $\rightarrow$ & ${\rm A^0 + g_{+}}$ \\
Molecular-ion attachment onto grains: & ${\rm HCO^{+} + g_{0}}$ & $\rightarrow$ & ${\rm HCO + g_{+}}$ \\
                                  & ${\rm HCO^{+} + g_{-}}$ & $\rightarrow$ & ${\rm HCO + g_{0}}$  \\
Charge transfer by grain-grain collisions:& ${\rm g_{+} + g_{-}}$ & $\rightarrow$ & ${\rm 2g_{0}}$\\
\tableline
\end{tabular}
\end{center}
\end{table}


\clearpage


\begin{thebibliography}{}

\bibitem[Abramowitz \& Stegun 1965]{AS}
Abramowitz, M. \& Stegun, I. 1965, {\it Handbook of Mathematical Functions}
(New York: Dover)  

\bibitem[Andre \etal 1993]{Andre}
Andr\'{e}, P., Ward-Tompson, D. \& Barsony, M. 1993, \apj, 406, 122 

\bibitem[Andre \etal 1996]{Andre96}
Andr\'{e}, P., Ward-Tompson, D. \& Motte, F. 1996, A\&A, 314, 625 

\bibitem[Arons \& Max 1975]{Arons}
Arons, J. \& Max, C. E. 1975, \apj, 196, L77

\bibitem[Bacmann \etal 2000]{Bacmann2000}
Bacmann, A., Andr\'{e}, P., Puget, J.-L., Abergel, A., Bontemps, S.,
\& Ward-Tompson, D. 2000, A\&A, 361, 555 

\bibitem[Basri \etal 1992]{Basri}
Basri, G., Marcy, G. W. \& Valenti, J. A. 1992, \apj, 390, 622 

\bibitem[Basu \& Mouschovias 1994]{BM94}
Basu, S., \& Mouschovias, T. Ch. 1994, \apj, 432, 720

\bibitem[Basu \& Mouschovias 1995]{BM95a}
Basu, S., \& Mouschovias, T. Ch. 1995a, \apj, 452, 386

\bibitem[Basu \& Mouschovias 1995]{BM95b}
Basu, S., \& Mouschovias, T. Ch. 1995b, \apj, 453, 271 

\bibitem[Beckwith \etal 1990]{Beckwith}
Beckwith, S. V. W., Sargent, A. I., Chini, R. S. \& Gusten, R. 1990, AJ, 
99, 924 

\bibitem[Benson \etal 1998]{Benson98}
Benson, P. J., Caselli, P., \& Myers, P. C. 1998, \apj, 
506, 743 

\bibitem[Bergin \etal 1999]{Bergin}
Bergin, E. A., Plume, R., Williams, J. P. \& Myers, P. C. 1999, \apj,
 512, 724 

\bibitem[Bonnor 1956]{Bonnor}
Bonnor, W. B. 1956, \mnras, 116, 351

\bibitem[Boss \& Black 1982]{BossBlack82}
Boss, A. P., \& Black, D. C. 1982, \apj, 258, 270

\bibitem[Carr 1987]{Carr87}
Carr, J. S. 1987, \apj, 323, 170 

\bibitem[Caselli \etal 1998]{Caselli}
Caselli, P., Walmsley, C. M., Tervieza, R. \& Herbst, E. 1998, \apj, 499, 234 

\bibitem[Ciolek \& Basu 2000]{CB2000}
Ciolek, G. E. \& Basu, S. 2000, \apj, 529, 925

\bibitem[Ciolek \& K\"{o}nigl 1998]{CK98}
Ciolek, G. E. \& K\"{o}nigl, A. 1998, \apj, 504, 257

\bibitem[Ciolek \& Mouschovias 1993]{Ciolek93}
Ciolek, G. E. \& Mouschovias, T. Ch. 1993, \apj, 418, 774


\bibitem{cm94}
Ciolek, G. E. \& Mouschovias, T. Ch. 1994, \apj, 425, 142

\bibitem[Ciolek \& Mouschovias 1995]{Ciolek95}
Ciolek, G. E. \& Mouschovias, T. Ch. 1995, \apj, 454, 194

\bibitem[Contopoulos, Ciolek \& K\"{o}nigl 1998]{CCK98}
Contopoulos, I., Ciolek, G. E. \& K\"{o}nigl, A. 1998, \apj, 504, 247

\bibitem[Crutcher 1999]{Crut99}
Crutcher, R. M. 1999, \apj, 520, 706

\bibitem[Crutcher \etal 1987]{Crutetal}
Crutcher, R. M., Kazes, I., Troland, T. H. 1987, A\&A, 181, 119  

\bibitem[Crutcher \etal 1994]{Crut94}
Crutcher, R. M., Mouschovias, T. Ch., Troland, T. H. \& Ciolek, G. E. 1994,
\apj, 427, 839

\bibitem[Crutcher \etal 1996]{Crut96}
Crutcher, R. M., Roberts, D. A., Mehringer, D. M., \& Troland, T. H. 1996,
\apj, 462, L79

\bibitem[Crutcher \etal 1993]{Crut93}
Crutcher, R. M., Troland, T. H., Goodman, A. A., Heiles, C.,
Kaz\'{e}s, I., \& Myers, P. C. 1993,
\apj, 407, 175


\bibitem[Desch \& Mouschovias 2001]{DM01}
Desch, S. J. \& Mouschovias, T. Ch. 2001, \apj, 550, 314

\bibitem[Draine \& Sutin 1987]{DraineSutin87}
Draine, B. T. \& Sutin, B. 1987, \apj, 320, 803

\bibitem[Ebert 1955]{Ebert1}
Ebert, R. 1955, Zs. Ap., 37, 217

\bibitem[Ebert 1957]{Ebert2}
Ebert, R. 1957, Zs. Ap., 42, 263

\bibitem[Eng \& Mouschovias 2004]{EM04}
Eng, C., \& Mouschovias, T. Ch. 2004, in preparation

\bibitem[]{fm92}
Fiedler, R. A. \& Mouschovias, T. Ch. 1992, \apj, 391, 199

\bibitem[]{fm93}
Fiedler, R. A. \& Mouschovias, T. Ch. 1993, \apj, 415, 680

\bibitem[Field 1970]{Field70}
Field, G. B. 1970, in {\it Interstellar Gas Dynamics},
ed. H. J. Habing (Dordrecht: Reidel), 51

\bibitem[Frerking \etal 1987]{Frerking} 
Frerking, M. A., Langer, W. D. \& Wilson, R. W. 1987, \apj, 313, 320

\bibitem[Galli \& Shu 1993a]{GS93a} 
Galli, D., \& Shu, F. H. 1993a, \apj, 417, 220

\bibitem[Galli \& Shu 1993b]{GS93b} 
Galli, D., \& Shu, F. H. 1993b, \apj, 417, 243

\bibitem[Goldsmith \& Arquilla 1985]{Goldsmith} 
Goldsmith, P. F. \& Arquilla, R. 1985, in {\it Protostars and Planets II},
ed. D. C. Black \& M. S. Matthews (Tucson: Univ. Arizona), 429 

\bibitem[Heiles \etal 1993]{Heiles93}
Heiles, C., Goodman, A., McKee, C. F., \& Zweibel, E. G. 1993, in {\it
  Protostars and Planets III},
ed. M. S. Matthews \& E. Levy (Tucson: Univ. Arizona Press), 279 

\bibitem[Hildebrand \etal 1999]{Hildeb}
Hildebrand, R. H., Dotson, J. L., Dowell, C. D., Schleuning, D. A. \&
Vaillancourt, J. E. 1999, \apj, 516, 834

\bibitem[]{khm84}
Kaifu, N., Hasegawa, T., Morimoto, M., Inatani, J., Nagane, K., Miyazawa, K.,
Chikada, Y., Kanzawa, T., Akabane, K., Suzuki, S. 1984, A\&A, 134, 7

\bibitem[Lai \etal 2002]{Lai}
Lai, S.-P., Crutcher, R. M., Girart, J. M., Rao, R. 2002, \apj, 566, 925

\bibitem[]{lar81}
Larson, R. B. 1981, \mnras, 194, 809

\bibitem[]{lkm82}
Leung, C. M., Kutner, M. L. \& Mead, K. N. 1982, \apj, 262, 583

\bibitem[]{lch94}
Lay, O. P., Carlstrom, J. E., Hills, R. E. \& Phillips, T. G. 1994, \apj, 
434, L75

\bibitem[Li 1998a]{li98a}
Li, Z.Y 1998a, ApJ, 493, 230

\bibitem[Li 1998b]{li98b}
Li, Z.Y 1998b, ApJ, 497, 850

\bibitem[]{lm96}
Li, Z.Y \& McKee C.F. 1996, ApJ, 464, 373

\bibitem[Li \& Shu]{ls97}
Li, Z.Y \& Shu F.H. 1997, \apj, 475, 237

\bibitem[Loren 1989]{Loren89}
Loren, R. B. 1989, ApJ, 338, 925

\bibitem[Mestel \& Spitzer 1956]{ms56}
Mestel, L. \& Spitzer, L., Jr. 1956, \mnras, 116, 503

\bibitem[]{mmc94}
Morton, S. A., Mouschovias, T. Ch. \& Ciolek, G. E. 1994, \apj, 421, 561

\bibitem[Mouschovias 1975]{TM75}
Mouschovias, T. Ch. 1975, PhD Thesis,  Univ. Calif., Berkeley

\bibitem[]{m76}
Mouschovias, T. Ch. 1976a, \apj, 206, 753

\bibitem[]{m76b}
Mouschovias, T. Ch. 1976b, \apj, 207, 141

\bibitem[]{ms76}
Mouschovias, T. Ch. \& Spitzer, L., Jr. 1976, \apj, 210, 326

\bibitem[]{m77}
Mouschovias, T. Ch. 1977, \apj, 211, 147

\bibitem[]{m78}
Mouschovias, T. Ch. 1978, in {\it Protostars and Planets}, ed. T. Gehrels
(Tucson: University of Arizona), 209

\bibitem[]{m79}
Mouschovias, T. Ch. 1979a, \apj, 228, 159

\bibitem[]{m79b}
Mouschovias, T. Ch. 1979b, \apj, 228, 475

\bibitem[]{m81}
Mouschovias, T. Ch. 1981, in {\it Fundamental Problems in the Theory of Stellar Evolution},
ed. D. Sugimoto, D. Q. Lamb, \& D. N. Schramm (Dordrecht: Reidel), 27

\bibitem[]{m87a}
Mouschovias, T. Ch. 1987a, in {\it Physical Processes in Interstellar Clouds},
ed. G. E. Morfill \& M. Scholer (Dordrecht: Reidel), 453

\bibitem[]{m87b}
Mouschovias, T. Ch. 1987b, in {\it Physical Processes in Interstellar Clouds},
ed. G. E. Morfill \& M. Scholer (Dordrecht: Reidel), 491

\bibitem[]{m89}
Mouschovias, T. Ch. 1989, in {\it The Physics and Chemistry of Interstellar
  Molecular Clouds},
ed. G. Winnewisser \& J. T. Armstrong (Berlin: Springer), 297


\bibitem[]{m91}
Mouschovias, T. Ch. 1991, \apj, 373, 169


\bibitem[]{m95}
Mouschovias, T. Ch. 1995, in {\it The Physics of the Interstellar Medium and 
Intergalactic Medium}, ed. A. Ferrara, C. F. McKee, C.Heiles \& P. R. Shapiro
(San Francisco: ASP), 80, 184

\bibitem[Mouschovias 1996]{m96}
Mouschovias, T. Ch. 1996, in {\it Solar and Astrophysical
  Magnetohydrodynamic Flows}, ed. K. Tsiganos (Dordrecht: Kluwer), 505


\bibitem[Mouschovias \& Morton 1985]{mm85a}
Mouschovias, T. Ch. \& Morton, S. A. 1985a, \apj, 298, 190

\bibitem[Mouschovias \& Morton 1985]{mm85b}
Mouschovias, T. Ch. \& Morton, S. A. 1985b, \apj, 298, 205

\bibitem[]{mm91}
Mouschovias, T. Ch. \& Morton, S. A. 1991, \apj, 371, 296

\bibitem[]{mm92a}
Mouschovias, T. Ch. \& Morton, S. A. 1992a, \apj, 390, 144

\bibitem[]{mm92b}
Mouschovias, T. Ch. \& Morton, S. A. 1992b, \apj, 390, 166

\bibitem[Mouschovias \& Paleologou 1979]{mp79}
Mouschovias, T. Ch. \& Paleologou, E. V. 1979, \apj, 
230, 204

\bibitem[Mouschovias \& Paleologou 1980]{mp80}
Mouschovias, T. Ch. \& Paleologou, E. V. 1980, \apj, 
237, 877 

\bibitem[]{mpf85}
Mouschovias, T. Ch., Paleologou, E. V. \& Fiedler, R. A. 1985, \apj, 
291, 772 

\bibitem[]{mps95}
Mouschovias, T. Ch. \& Psaltis, D. 1995, \apj, 444, L105 

\bibitem[]{my83}
Myers, P. C. 1983, \apj, 270, 105 

\bibitem[]{myb83}
Myers, P. C. \& Benson, P. J. 1983, \apj, 266, 309 

\bibitem[]{mylb83}
Myers, P. C., Linke, R. A. \& Benson, P. J. 1983, \apj, 264, 517 

\bibitem[]{nn78}
Nakano, T. \& Nakamura, T. 1978, PASJ, 30, 671

\bibitem[]{nnu91}
Nishi, R., Nakano, T. \& Umebayashi, T. 1991, \apj, 368, 181

\bibitem[]{pm83}
Paleologou, E. V. \& Mouschovias, T. Ch. 1983, \apj, 275, 838

\bibitem[]{pm65}
Pneuman, G. W. \& Mitchell, T. P. 1965, {\it Icarus}, 4, 494

\bibitem[Rao \etal 1998]{Rao}
Rao, R., Crutcher, R. M., Plambeck, R. L. \& Wright, M. C. H. 1998, \apjl,
502, L75

\bibitem[Safier \etal 1997]{SMS97}
Safier, P. N., McKee, C. F., \& Stahler, S. W. 1997, \apj, 485, 660


\bibitem[Saito \etal 1996]{Saito}
Saito, M., Kawabe, R., Kitamura, Y. \& Sunada, K. 1996, \apj, 453, 384

\bibitem[]{}
Sargent, A. I., Beckwith, S., Keene, J. \& Masson, C. R. 1988, \apj, 
333, 936
 
\bibitem[Shu 1977]{Shu77}
Shu, F. H. 1977, \apj, 214, 488

\bibitem[Shu \etal 1999]{Shu99}
Shu, F. H., Allen, A., Shang, H., Ostriker, E. C., \& Li, Z.-Y. 1999, 
in {\it The Origin of Stars and Planetary Systems}, ed. Lada,
C. J. \& Kylafis, N. D. (Dordrechet: Kluwer), 193

\bibitem[]{sp41}
Spitzer, L. Jr. 1941, \apj, 93, 369

\bibitem[]{sp47}
Spitzer, L. Jr. 1947, \apj, 107, 6

\bibitem[Tafalla \etal 1998]{tafalla98}
Tafalla, M., Mardones, D., Myers, P.C., Caselli, P., Bachiller, R, \&
Benson, P. J. 1998, \apj, 504, 900 

\bibitem[]{paperII}
Tassis, K. \& Mouschovias, T. Ch. 2004b, ApJ {\em submitted}

\bibitem[]{MCages}
Tassis, K. \& Mouschovias, T. Ch. 2004c, ApJ, in print

\bibitem[Tomisaka 1996]{T96}
Tomisaka, K. 1996, PASJ, 48, L97

\bibitem[Tscharnuter \& Winkler 1979]{TW79}
Tscharnuter, W. M. \& Winkler, K. H. 1979, Pub. Astr. Soc. Japan, 32, 405

\bibitem[Umebayashi \& Nakano 1980]{UN80}
Umebayashi, T. \& Nakano, T. 1980, Comput. Phys. Comm., 18, 171

\bibitem[van Leer 1979]{vL79}
Van Leer, B. 1979, J. Comput. Phys., 32, 101

\bibitem[Ward-Thompson \etal 1994]{WT94}
Ward-Thompson, D., Scott, P. F., Hills, R. E., \& Andr\'{e}, P. 1994,
MNRAS, 268, 276

\bibitem[Wade \etal 1998]{Wade}
Wade, G. A., Hill, G. M., Adelman, S. J., Manset, N. \& Bastien, P. 1998,
A\&A, 335, 973

\bibitem[Williams \etal 1998]{Williams}
Williams, J. P., Bergin, E. A., Caselli, P., Myers, P. C. \& Plume, R. 1998,
\apj, 503, 689

\bibitem[Williams \etal 1999]{W99}
Williams, J. P., Myers, P. C., Wilner, D. J., \& Di Francesco, J. 1999,
\apj, 513, L61


\bibitem[Zhou \etal 1990]{Zhou}
Zhou, S., Evans, N. J., II, Butner, H. M., Kutner, M. L., Leung, C. M. \&
Moundy, L. G. 1990, \apj, 363, 168



\end{thebibliography}
\end{document}